\documentclass[12pt,preprintnumbers,eqsecnum]{revtex4}

\usepackage{amsfonts}
\usepackage{mathrsfs}
\usepackage{amssymb}
\usepackage{amsmath}
\usepackage{graphicx,epsfig}
\usepackage{pstricks}
\usepackage{slashed}

\def\nn{\nonumber}
\def\be{\begin{equation}}
\def\ee{\end{equation}}
\def\bea{\begin{eqnarray}}
\def\eea{\end{eqnarray}}
\def\bean{\begin{eqnarray*}}
\def\eean{\end{eqnarray*}}
\def\bary{\begin{array}}
\def\eary{\end{array}}
\def\bit{\begin{itemize}}
\def\eit{\end{itemize}}

\def\su5u1{SU(5) \times U(1)}
\def\fsu5u1{SU(5) \times U(1)'}
\def\so10{SO(10)}
\def\sq20{SO(10) \times SO(10)}

\def\bwt{\begin{widetext}}
\def\ewt{\end{widetext}}
\def\be{\begin{equation}}
\def\ee{\end{equation}}
\def\bea{\begin{eqnarray}}
\def\eea{\end{eqnarray}}
\def\bean{\begin{eqnarray*}}
\def\eean{\end{eqnarray*}}
\def\bary{\begin{array}}
\def\eary{\end{array}}
\def\bit{\begin{itemize}}
\def\eit{\end{itemize}}

\def\su5u1{SU(5) \times U(1)}
\def\fsu5u1{SU(5) \times U(1)'}
\def\so10{SO(10)}
\def\sq20{SO(10) \times SO(10)}

\usepackage{amssymb}

\begin{document}

\setlength{\parskip}{0.1cm}

\preprint{MIFPA-13-10,\, OSU-HEP-13-01, \, RECAPP-HRI-2013-004}

\title{\Large Top $SU(5)$ Models: Baryon and Lepton Number Violating Resonances at the LHC}
\author{Shreyashi Chakdar$^{1,}$\footnote{Electronic address: chakdar@okstate.edu},
Tianjun Li$^{2,3,}$\footnote{Electronic address: tli@itp.ac.cn},
S. Nandi$^{1,}$\footnote{Electronic address: s.nandi@okstate.edu}, and
Santosh Kumar Rai$^{4,}$\footnote{Electronic address: skrai@hri.res.in}}
\affiliation{$^1$Department of Physics and Oklahoma Center for High Energy Physics, 
Oklahoma State University, Stillwater OK 74078-3072, USA. \\
$^2$State Key Laboratory of Theoretical Physics 
and Kavli Institute for Theoretical Physics China (KITPC), 
      Institute of Theoretical Physics, Chinese Academy of Sciences, 
Beijing 100190, P. R. China. \\
$^3$George P. and Cynthia W. Mitchell Institute for
Fundamental Physics and Astronomy,   Texas A$\&$M University,
College Station, TX 77843, USA. \\
$^4$Regional Centre for Accelerator-based Particle Physics,  Harish-Chandra Research Institute,  \\ Chhatnag Road, Jhusi, 
Allahabad 211019, India.}
%

\begin{abstract}
We propose the minimal and renormalizable non-supersymmetric top $SU(5)$ models where the $SU(5)\times SU(3)'_C \times SU(2)'_L \times U(1)'_Y$ gauge symmetry is broken down to the Standard Model (SM)
gauge symmetry 
at the TeV scale. The first 
two families of the SM fermions are charged under $ SU(3)'_C \times SU(2)'_L \times U(1)'_Y$ while the third family is charged under $SU(5)$. 
In the minimal top $SU(5)$ model, we show that the quark CKM mixing
matrix can be generated via dimension-five operators, and the
proton decay problem can be solved by fine-tuning the coefficients of the
higher dimensional operators at the order of $10^{-4}$.
In the renormalizable top $SU(5)$ model, we can explain 
the quark CKM mixing matrix by introducing vector-like particles,
and we do not have proton decay problem. The models give rise to leptoquark and diquark gauge bosons which violate both lepton and baryon numbers involving the third family quarks and leptons. The current experimental limits for these particles is well below the TeV scale. We also discuss the productions and decays of these new gauge bosons, and their ensuing signals, as well as their reach at the LHC. 

\end{abstract}



\maketitle

\newpage

\section{Introduction}

 The Standard Model (SM), based on the local
gauge symmetry $SU(3)_C\times SU(2)_L\times U(1)_Y$, is very
successful in describing all the experimental results below
the  TeV scale. It is an excellent effective field theory, but
it is widely believed not to be  the final theory.  
 Discovery of new particles is  highly anticipated at 
the Large Hadron Collider (LHC). The most likely and reasonably
well motivated candidates are  supersymmetric particles, 
and extra $Z'$ boson. However, it is important to explore other alternatives or entirely new possibilities at the current and future LHC.

In the SM, we have fermions (spin 1/2) and scalars 
(Higgs fields)(spin 0) which do not belong to adjoint 
representations under the SM gauge symmetry. 
Can we also have TeV scale gauge bosons (spin 1) belonging
to the non-adjoint representations under the SM 
gauge symmetry? Can we achieve the (partial) grand unified theory at the TeV scale? Can we construct a renormalizable theory realizing such a possibility which can be  tested at the LHC? These are very interesting theoretical questions that we shall address in this work. Discovery of such gauge  bosons around the TeV scale at the LHC will open up a new window for our understanding of the fundamental theory describing the nature.

How can we construct a consistent  theory involving the massive vector bosons which do not belong to the adjoint representations under the SM gauge symmetry? If the massive vector bosons are not the gauge bosons of a symmetry group, there are some theoretical problems from the consistency of quantum field theory, for instance, the unitarity and renormalizability~\cite{Lee:1962vm}.
When the gauge symmetry is spontaneously
broken via the Higgs mechanism, the interactions of
the massive gauge bosons satisfy both the unitarity and the
renormalizability of the theory~\cite{Hooft,Lee}.
Thus, the massive vector bosons must be the gauge bosons arising from  spontaneous gauge symmetry breaking.

As we know, a lot of models with extra TeV scale gauge bosons have been proposed previously in the literature. However, those massive gauge bosons  either  belong to the adjoint representations or are singlets under the SM gauge symmetry~\cite{Hill:1991at,Hill:1993hs,Dicus:1994sw,
Muller:1996dj,Malkawi:1996fs,Erler:2002pr,Chiang:2007sf,PSLR}. For example, in the top color model~\cite{Hill:1991at,Hill:1993hs,Dicus:1994sw}, 
the colorons belong to the adjoint representation
of the $SU(3)_C$; in the top 
flavor model~\cite{Muller:1996dj,Malkawi:1996fs}, the extra $W'$ and $Z'$
bosons belong to the adjoint representation of the $SU(2)_L$, while
in the $U(1)'$ model~\cite{Erler:2002pr} or 
top hypercharge model~\cite{Chiang:2007sf}, 
the new $Z'$ boson is a singlet under the SM gauge symmetry.
In the Grand Unified Theories such as $SU(5)$ and
 $SO(10)$~\cite{Georgi:1974sy,Fritzsch:1974nn},
there are such kind of massive gauge bosons. However, their
masses have to be around the unification scale $\sim 10^{16}$ GeV 
to satisfy the proton decay constraints. 

Some years ago, two of us (TL and SN) had  proposed
 a class of models where the gauge symmetry is  ${\cal G} \equiv 
\prod_i G_i \times SU(3)'_C \times SU(2)'_L \times U(1)'_Y$~\cite{Li:2004cj}.
The quantum numbers of the SM fermions and Higgs fields under
the $SU(3)'_C \times SU(2)'_L \times U(1)'_Y$ gauge symmetry are
the same as they have under the SM gauge 
symmetry $SU(3)_C \times SU(2)_L \times U(1)_Y$,
while they are all singlets under $\prod_i G_i$. Hence $\prod_i G_i$ is 
the hidden gauge symmetry. After the 
gauge symmetry ${\cal G}$ is spontaneously broken down to
the SM gauge symmetry at the TeV scale via Higgs mechanism,
some of the massive gauge bosons from the ${\cal G}$ breaking
do not belong to the adjoint representations under the SM
gauge symmetry. In particular, a
 concrete $SU(5) \times SU(3)'_C \times SU(2)'_L \times U(1)'_Y$
has been studied in detail.
However, the corresponding $(X_{\mu}, Y_{\mu})$ massive gauge 
bosons are meta-stable and behave like the stable heavy
quarks and anti-quarks at the LHC~\cite{Li:2004cj}.
Thus, an interesting question is whether we can construct the
$SU(5) \times SU(3)'_C \times SU(2)'_L \times U(1)'_Y$
models where the $(X_{\mu}, Y_{\mu})$ gauge bosons can decay and produce interesting signals at the LHC. By the way, the six-dimensional orbifold non-supersymmetric and supersymmetric $SU(5)$ and $SU(6)$ models with low energy gauge unification
have been constructed previously~\cite{Li:2001qs, Li:2002xw, Jiang:2002at}. However, there is no direct interactions between the  $(X_{\mu}, Y_{\mu})$ particles and the SM fermions.

As we pointed out above, the top color model~\cite{Hill:1991at,Hill:1993hs,Dicus:1994sw}, 
top flavor model~\cite{Muller:1996dj,Malkawi:1996fs},
and top hypercharge model~\cite{Chiang:2007sf} have been constructed before. 
Because of the proton decay problem and quark CKM mixings,  etc,
the real challenging question is whether we can construct the top $SU(5)$ model as the unification of these models. Consequently,
we can explain the charge quantization for the third family, and 
 probe  the baryon and lepton number violating interactions involving the third family at the LHC. Such a model was proposed by us recently \cite{Chakdar:2012kd}, and its implications for LHC was briefly explored.  

In this paper, we shall propose  two such models: the minimal and the renormalizable top $SU(5)$ model where the 
$SU(5)\times SU(3)'_C \times SU(2)'_L \times U(1)'_Y$ gauge symmetry 
is broken down to the SM
gauge symmetry via the bifundamental Higgs fields at low energy. The first 
two families of the SM fermions are charged under $ SU(3)'_C \times SU(2)'_L \times U(1)'_Y$
while the third family is charged under $SU(5)$. 
In the minimal top $SU(5)$ model, we show that the quark CKM mixing matrix can be generated via dimension-five operators, and the proton decay problem can be solved by fine-tuning the coefficients of the higher dimensional operators at the order of $10^{-4}$. In the renormalizable top $SU(5)$ model, we can explain 
the quark CKM mixing matrix by introducing vector-like particles, and we do not have proton decay problem. In these models, the non-unification of the three SM couplings are remedied, because three SM couplings $g_3$, $g_2$, $g_1$ are now combinations of $(g_5, g_3')$, $(g_5, g_2')$,$(g_5, g_1')$, and need not be unified. Since the models have baryon and lepton number violating interactions, it might be useful in generating the baryon asymmetry of the Universe. In our models, since the third family quark lepton unification is at the TeV scale,  we can probe the new $(X_{\mu}, ~Y_{\mu})$ gauge bosons at the LHC through their decays  to the third family of the SM fermions. 

Our paper is organized as follows. In section II, we discuss the two models and their formalism. In section III, we discuss in detail the phenomenological implications of the models. These include the productions and decays of the X and Y gauge bosons at the LHC energies of 7, 8 and 14 TeV, their  decay modes, and the signals for the final states. We also discuss the LHC reach for the masses of these particle for various LHC energies and luminosities. Section IV contains our summary and conclusions.

\section{The Minimal and Renormalizable Top $SU(5)$ Models}

We propose two non-supersymmetric top $SU(5)$ models 
where the gauge symmetry is 
$ SU(5)\times SU(3)'_C \times SU(2)'_L \times U(1)'_Y$.
The first 
two families of the SM fermions are charged under 
$ SU(3)'_C \times SU(2)'_L \times U(1)'_Y$
while the third family is charged under $SU(5)$. 
We denote the gauge fields
for $SU(5)$ and $SU(3)'_C \times SU(2)'_L \times U(1)'_Y $
as ${\widehat A}_{\mu}$ and ${\widetilde A}_{\mu}$, respectively,
and the gauge couplings for $SU(5)$, $SU(3)'_C$, $SU(2)'_L$
and $U(1)'_Y$ are $g_5$, $g'_3$, $g'_2$ and $g'_Y$,
respectively. The Lie algebra indices for the generators of
 $SU(3)$, $SU(2)$ and $U(1)$
are denoted by $a3$, $a2$ and $a1$, respectively, and
the  Lie algebra indices for the generators of
$SU(5)/(SU(3)\times SU(2)\times U(1))$ are denoted by ${\hat a}$.
After the $SU(5)\times SU(3)'_C \times SU(2)'_L \times U(1)'_Y $
gauge symmetry is broken down to the SM gauge symmetry
$SU(3)_C\times SU(2)_L \times U(1)_Y$,
 we denote the massless gauge fields for the
SM gauge symmetry as $A_{\mu}^{ai}$, and
the massive gauge fields as $B_{\mu}^{ai}$ and ${\widehat A}_{\mu}^{\hat a}$.
The gauge couplings for the SM gauge symmetry
$SU(3)_C$, $SU(2)_L$ and $U(1)_Y$ are $g_3$, $g_2$ and $g_Y$, respectively.

To break the $SU(5)\times SU(3)'_C \times SU(2)'_L \times U(1)'_Y $
gauge symmetry down to the SM gauge symmetry, we introduce
two bifundamental Higgs fields $U_T$ and $U_D$~\cite{Li:2004cj}. 
Let us explain our convention. We denote the first two family
quark doublets, right-handed up-type quarks, right-handed down-type 
quarks, lepton doublets, right-handed neutrinos, right-handed charged leptons,
and the corresponding Higgs field respectively
as $Q_i$, $U_i^c$, $D_i^c$, $L_i$, $N^c_i$, $E_i^c$, and $H$, as in the 
supersymmetric SM convention. We denote the third family 
SM fermions as $F_3$, $\overline{f}_3$, and $N_3^c$.
To give the masses to the third family of the SM fermions, we introduce
a $SU(5)$ anti-fundamental Higgs field $\Phi\equiv (H'_T, H')$. We also
 need to introduce a scalar field $XT$ if we require that the triplet
Higgs $H'_T$ have mass around the 
$SU(5)\times SU(3)'_C \times SU(2)'_L \times U(1)'_Y$ gauge symmetry breaking
scale. However, it is not necessary, and we will explain it in
the following. In addition, note that the neutrino PMNS mixings can be
generated via the right-handed neutrino Majorana mass mixings.
we propose two top $SU(5)$ models which can generate the mass for the possible
pseudo-Nambu-Goldston boson (PNGB) $\phi$ during the gauge symmetry
breaking and generate the quark CKM mixings.
In the minimal top $SU(5)$ model,  we consider the dimension-five
non-renormalizable operators and fine-tune some
coefficients of the higher dimensional operators
at the order $10^{-4}$ to suppress the proton decay.
In the renormalizable top $SU(5)$ model,  we introduce the additional vector-like 
particles. To give the PNGB mass, we introduce a scalar field $XU$ in the
$SU(5)$ anti-symmetric representation.
And to generate the quark CKM mixings while not to introduce
the proton decay problem, we introduce the vector-like fermionic particles 
$(Xf, ~Xf^c)$ and $(XD, ~XD^c)$. Note that the 
$SU(3)'_C \times SU(2)'_L \times U(1)'_Y $ gauge symmetry can be formally embedded
into a global $SU(5)'$ symmetry, and to do that, we introduce the vector-like particles
$(XL, ~XL^c)$ as well. The complete
particle content and the particle quantum numbers under 
$SU(5)\times SU(3)'_C \times SU(2)'_L \times U(1)'_Y $ gauge symmetry
are given in Table~\ref{tab:Content}.

\begin{table}[htb]
\caption[]{The complete
particle content and the particle quantum numbers under 
$SU(5)\times SU(3)'_C \times SU(2)'_L \times U(1)'_Y $ gauge symmetry in the
top $SU(5)$ model. Here, $i=1,~2$, and $k=1,~2,~3$.}
\label{tab:Content}
\begin{center}
  \begin{tabular}{|c|c||c|c|}
   \hline
 Particles  & Quantum Numbers  & Particles & Quantum Numbers \\ \hline
 $Q_i$ & $({\bf 1}; {\bf {3}}, {\bf 2}, {\bf {1/6}})$  
& $L_i$ & $({\bf 1}; {\bf {1}}, {\bf 2}, {\bf {-1/2}})$  
\\\hline
 $U^c_i$ & $({\bf 1}; {\bf {\bar 3}}, {\bf 1}, {\bf {-2/3}})$  
& $N^c_k$ & $({\bf 1}; {\bf {1}}, {\bf 1}, {\bf {0}})$  
\\\hline
 $D^c_i$ & $({\bf 1}; {\bf {\bar 3}}, {\bf 1}, {\bf {1/3}})$  
& $E^c_i$ & $({\bf 1}; {\bf {1}}, {\bf 1}, {\bf {1}})$  
\\\hline
 $F_3$ & $({\bf 10}; {\bf {1}}, {\bf 1}, {\bf {0}})$  
& $\overline{f}_3$ & $({\bf {\bar 5}}; {\bf {1}}, {\bf 1}, {\bf {0}})$  
\\\hline
 $H$ & $({\bf 1}; {\bf {1}}, {\bf 2}, {\bf {-1/2}})$  
& $\Phi$ & $({\bf {\bar 5}}; {\bf {1}}, {\bf 1}, {\bf {0}})$  
\\\hline
 $U_T$ & $({\bf 5}; {\bf {\bar 3}}, {\bf 1}, {\bf {1/3}})$
& $U_D$ & $({\bf 5}; {\bf {1}}, {\bf { 2}}, {\bf -1/2})$
\\\hline \hline
$XT$ & $({\bf 1}; {\bf {\bar 3}}, {\bf 1}, {\bf {1/3}})$ &
$XU$ & $({\bf 10}; {\bf {1}}, {\bf 1}, {\bf {-1}})$ 
\\\hline 
 $Xf$ & $({\bf 5}; {\bf {1}}, {\bf 1}, {\bf {0}})$  
& $\overline{Xf}$ & $({\bf {\bar 5}}; {\bf {1}}, {\bf 1}, {\bf {0}})$  
\\\hline
$XD$ & $({\bf 1}; {\bf {3}}, {\bf 1}, {\bf {-1/3}})$  
& $\overline{XD}$ & $({\bf {1}}; {\bf {\bar 3}}, {\bf 1}, {\bf {1/3}})$  
\\\hline
$XL$ & $({\bf 1}; {\bf {1}}, {\bf 2}, {\bf {-1/2}})$  
& $\overline{XL}$ & $({\bf {1}}; {\bf {1}}, {\bf 2}, {\bf {1/2}})$  
\\\hline
\end{tabular}
\end{center}
\end{table}

To give the vacuum expectation values (VEVs)
 to the bifundamental Higgs fields $U_T$ and $U_D$, we consider the
following Higgs potential
\begin{eqnarray}
V &=& -m_{T}^2 |U_T^2|-m_{D}^2 |U_D^2| + \lambda_T |U_T^2|^2
+ \lambda_D |U_D^2|^2 + \lambda_{TD} |U_T^2| |U_D^2|
\nonumber\\ &&
+\left[ A_T \Phi U_T XT^{\dagger} + A_D \Phi U_D H^{\dagger} +
 {{y_{TD}}\over {M_{*}}} U^3_T U^2_D + {\rm H.C.}\right]~,~\,
\label{potential}
\end{eqnarray} 
where $M_*$ is a normalization mass scale.

A few remarks are in order. First, with $XT$ particle, 
the Higgs triplet $H'_T$ will have mass around the 
$SU(5) \times SU(3)'_C \times SU(2)'_L \times U(1)'_Y$
gauge symmetry breaking scale, as given by the above
$A_T$ term. However, it is still fine even if we do not
introduce the $XT$ field. Let us explain it in detail.
In our models, we have
two Higgs doublets $H$ and $H'$, which give
the masses to the first two families and
the third family of the SM fermions, respectively. Thus,
$H'_T$ will have mass around a few hundred GeV,
and it has interesting decay channels via Yukawa couplings, 
which will be discussed in the following.

Second, without the non-renormalizable $y_{TD}$ term,
we have global symmetry $U(5)\times SU(3)'_C \times SU(2)'_L \times U(1)'_Y $
in the above potential, and then we will have a PNGB $\phi$
during the $SU(5) \times SU(3)'_C \times SU(2)'_L \times U(1)'_Y$ gauge
symmetry breaking. To break the $U(5)$ global symmetry
down to $SU(5)$ and then give mass to $\phi$, we do need this non-renormalizable term.
Moreover, $M_*$ can be around the intermediate scale, for example,
1000~TeV. If we assume that all the high-dimensional operators are 
suppressed by the reduced Planck scale, {\it i.e.}, $ M_*=M_{\rm Pl}$, 
we can generate the $y_{TD}$ term by introducing the $XU$ field. 
The relevant Lagrangian is 
\begin{eqnarray}
-{\cal L} &=& \left(y_T U_T^3 XU + y_D \mu' U_D^2 XU^{\dagger} + {\rm H.C.} \right)
+ M^2_{XU} |XU|^2~,~\, 
\label{Eq-P-XU}
\end{eqnarray} 
where the mass scales
$\mu'$ and  $M_{XU} $ will be assumed to be around 1000~TeV. 
After we integrate
out $XU$, we get the needed high-dimensional operator
\begin{eqnarray}
V \supset- {{y_T y_D \mu'}\over {M_{XU}^2}}  U^3_T U^2_D~.~\,
\end{eqnarray} 

We choose the following VEVs for the fields $U_T$ and $U_D$
\begin{eqnarray}
 <U_T> =  {v_T} \left(
  \begin{array}{c}
    I_{3\times3} \\
    0_{2\times3} \\
  \end{array}
  \right)~, \quad
 <U_D> =  {v_D} \left(
  \begin{array}{c}
    0_{3\times2} \\
    I_{2\times2} \\
  \end{array}
  \right)~,
\end{eqnarray}
where $I_{i\times i}$ is the $i\times i$ identity matrix, and
$0_{i\times j}$ is the $i\times j$ matrix where all the entries are
zero. We assume that $v_D$ and $v_T$ are in the 
TeV range so that the massive gauge bosons  have TeV scale masses.

From the kinetic terms for the fields
$U_T$ and $U_D$ , we obtain the mass terms for the gauge fields 
\begin{eqnarray}
\sum_{i=T, D} \langle (D_{\mu} U_i)^{\dagger} D^{\mu} U_i \rangle
&=& {1\over 2} v_T^2   \left( g_5 {\widehat A}_{\mu}^{a3} - 
g'_3 {\widetilde A}_{\mu}^{a3} \right)^2
+ {1\over 2} v_D^2  \left( g_5 {\widehat A}_{\mu}^{a2} - 
g'_2 {\widetilde A}_{\mu}^{a2} \right)^2
\nonumber\\ &&
+\left( {{v_T^2}\over 3} 
+ {{v_D^2}\over 2}  \right) 
\left( g_5^Y {\widehat A}_{\mu}^{a1} - 
g'_Y {\widetilde A}_{\mu}^{a1} \right)^2
\nonumber\\ &&
+ {1\over 2} g_5^2 \left(v_T^2  +v_D^2 \right) 
\left(X_{\mu} \overline{X}_{\mu}
+ Y_{\mu} \overline{Y}_{\mu}\right)
~,~\,
\label{massterm}
\end{eqnarray}
 where $g_5^Y \equiv {\sqrt 3} g_5/{\sqrt 5}$,
 and we define the complex fields
($X_{\mu}$, $Y_{\mu}$) 
with quantum numbers (${\bf 3}$, ${\bf 2}$, ${\bf {5/6}}$) 
from the gauge fields ${\widehat A}_{\mu}^{\hat a}$, similar to that
in the usual $SU(5)$ model~\cite{Georgi:1974sy}.

From the original gauge fields ${\widehat A}_{\mu}^{ai}$ and
${\widetilde A}_{\mu}^{ai}$ and from
Eq. (\ref{massterm}), we obtain the massless gauge bosons $A_{\mu}^{ai}$ and the
TeV scale massive gauge bosons $B_{\mu}^{ai}$ ($i=3, 2, 1$)
which are in the adjoint representations of the SM gauge
symmetry
\begin{eqnarray}
\left(
\begin{array}{c}
A_\mu^{ai} \\
B_\mu^{ai}
\end{array} \right)=
\left(
\begin{array}{cc}
\cos\theta_i & \sin\theta_i \\
-\sin\theta_i & \cos\theta_i
\end{array}
\right)
\left(
\begin{array}{c}
{\widehat A}_\mu^{ai} \\
{\widetilde A}_\mu^{ai}
\end{array} \right)
~,~\,
\end{eqnarray}
where $i=3, 2, 1$, and 
\begin{eqnarray}
\sin\theta_j \equiv {{g_5}\over\displaystyle {\sqrt {g_5^2 +(g'_j)^2}}}
~,~
\sin\theta_1 \equiv {{g^Y_5}\over\displaystyle 
{\sqrt {(g_5^{Y})^2 +(g_Y^{\prime})^2}}} ~,~\,
\end{eqnarray}
where $j=3, 2$.
We also have the massive gauge bosons
($X_{\mu}$, $Y_{\mu}$) and (${\overline{X}_{\mu}}$, ${\overline{Y}_{\mu}}$)
which are not in the adjoint representations of the SM gauge symmetry.
 So, the $SU(5)\times SU(3)'_C \times SU(2)'_L \times U(1)'_Y $
gauge symmetry is  broken down to the diagonal SM gauge symmetry
$SU(3)_C\times SU(2)_L \times U(1)_Y$, and the theory is
unitary and renormalizable. The SM gauge couplings
$g_j$ ($j=3, 2$) and $g_Y$ are given by
\begin{eqnarray}
{1\over {g_j^2}} ~=~ {1\over {g_5^2}} + {1\over {(g'_j)^2}}~,~
{1\over {g_Y^2}} ~=~ {1\over {(g^Y_5)^2}} + {1\over {(g'_Y)^2}}~.~\,
\end{eqnarray}

If the theory is perturbative,  the upper and lower
bounds on the gauge couplings $g_5$, $g'_3$, $g'_2$ and $g'_Y$ are 
\begin{eqnarray}
&& g_3 ~< ~g_5 ~< ~{\sqrt {4\pi}} ~,~ g_3 ~<~ g'_3 ~<~ {\sqrt {4\pi}} ~,~
\\ &&
g_2 ~<~ g'_2 ~<~
{{g_3 g_2}\over {\sqrt {g_3^2 -g_2^2}}}~,~ \\ &&
g_Y ~<~ g'_Y  ~<~
{{{\sqrt 3} g_3 g_Y}\over {\sqrt {3g_3^2 -5g_Y^2}}}~.~ \,
\end{eqnarray}
Note that the gauge coupling  $g_5$ for $SU(5)$  is naturally 
large at the TeV scale because
the beta function of $SU(5)$ is negative, {\it i.e.},
$SU(5)$ is asymptotically free.

\subsection{The Minimal Model}

We consider the minimal model first, where we do not 
introduce any extra (``$X$'') particles $XT$, $XU$, $Xf$,
$\overline{Xf}$, $XD$, $\overline{XD}$, $XL$, and
$\overline{XL}$. So the Higgs triplet $H'_T$ will be 
a few hundred GeV.
We introduce the non-renormalizable
operators to generate the quark CKM mixings. We also escape the 
proton decay problem by fine-tuning some coefficients
of the higher-dimensional operators.

The renormalizable SM fermion Yukawa couplings are
\begin{eqnarray}
-{\cal L} &=& y^u_{ij} U_i^c Q_j {\widetilde H} +
y^{\nu}_{kj} N_k^c L_j  {\widetilde H} + 
 y^d_{ij} D_i^c Q_j H + y^e_{ij} E_i^c L_j H 
\nonumber\\ &&
+y^u_{33} F_3 F_3 \Phi^{\dagger}
+ y^{de}_{33} F_3 {\overline f}_3 \Phi
+ y^{\nu}_{k3} N_{k}^c  {\overline f}_3 \Phi^{\dagger}
+ m^N_{kl}  N_{k}^c  N_{l}^c  + {\rm H.C.} ~,~\,
\end{eqnarray}
where $i/j=1,~2$, $k/l=1,~2,~3$, and $ {\widetilde H}=i\sigma_2 H^{\dagger}$
with $\sigma_2$ the second Pauli matrix.
Because the three right-handed neutrinos can mix among
themselves via the Majorana masses, we can generate
the observed neutrino masses and mixings.
In addition, we make a wrong prediction that the bottom Yukawa coupling is
equal to the tau Yukawa coupling at the low energy. 
We can easily avoid this problem by introducing
the high-dimensional Higgs field under $SU(5)$, which
is out of the scope of this paper.
In addition, the Yukawa terms between 
the triplet Higgs field $H'_T$ in $\Phi$ and
 the third family of the SM fermions are
 $ y^{de}_{33} t^c b^c H'_T$, $ y^{de}_{33} Q_3 L_3 H'_T$, 
and $y^{u}_{33} t^c \tau^c H^{\prime \dagger}_T$. So, we have
$(B+L)$ violating interactions as well.

To generate the quark CKM mixings, we consider the higher-dimensional
operators. The dimension-five operators are
\begin{eqnarray}
-{\cal L} &=& {1\over {M_*}}
\left(  y^d_{i3} D_i^c F_3 \Phi U_T^{\dagger}
+ y^{e}_{i3} E_i^c {\overline f}_3 H U_D 
+
y^d_{3i} {\overline f}_3  Q_i H U_T
+ y^{e}_{3i} F_3 L_i \Phi U_D^{\dagger} \right) + {\rm H.C.} 
 ~.~\,
\label{O-Dim-5}
\end{eqnarray}
And the dimension-six operators are
\begin{eqnarray}
-{\cal L} &=&  {1\over {M^2_*}}
\left(y^u_{i3} U_i^c F_3 {\widetilde H} U_T^{\dagger} U_D^{\dagger}
+ y_{i3}^{\prime d} D_i^c F_3 H U_T^{\dagger} U_D^{\dagger}
+ y^u_{3i} F_3 Q_i \Phi^{\dagger} U_T U_D
\right.\nonumber\\ &&\left.
+ y^{\prime d}_{3i} \overline{f}_3 Q_i \Phi U_T U_D \right) + {\rm H.C.} 
 ~.~\,
\label{O-Dim-6}
\end{eqnarray}
Interestingly, if we neglect the dimension-six operators in Eq.~(\ref{O-Dim-6}),
we will generate the down-type quark mixings and charged
lepton mixings via the dimension-five operators in Eq.~(\ref{O-Dim-5}). 
Thus, the quark CKM mixing matrix can be
realized via the down-type quark mixings. 
The  proton decay is not a problem since there
is no mixing between the top quark and up quark.
For example, if we assume that the Yukawa couplings
$y^d_{i3}$ and $y^d_{3i}$ are order one and the VEVs of
$U_T$ and $U_D$ are about 1 TeV, we get $M_* \sim 1000$~TeV to generate the correct CKM mixings.

However, if we introduce the above dimension-six operators
in Eq.~(\ref{O-Dim-6}),
proton decay can indeed arises due to the up-type quark mixings.
For simplicity,  we assume that $y^d_{i3}$ and $y^d_{3i}$ are 
order one, $M_* \sim 1000$~TeV, and
the other Yukawa couplings $y^{e}_{i3}$, $y^{e}_{3i}$,
$y^u_{i3}$, $y^u_{3i}$ are very small and of the the same order.
Noting that the dimension-six proton decay operators have two up
quarks, one down quark and one lepton, 
from the current proton decay constraints, we obtain that
the Yukawa couplings $y^{e}_{i3}$, $y^{e}_{3i}$,
$y^u_{i3}$, $y^u_{3i}$ are about $10^{-4}$.
Because $m_e/m_t \sim 10^{-5}$, our fine-tuning is 
 one order smaller and therefore is still acceptable. We would like to point
out that the tau lepton decays to electron and muon will
be highly suppressed due to the very small $y^{e}_{i3}$ and $y^{e}_{3i}$
in the minimal model.
 
\subsection{The Renormalizable Model}

In the renormalizable model, we assume that all the non-renormalizable
operators are suppressed by the reduced Planck scale.
Thus, we need to introduce all the particles in Table~\ref{tab:Content}.
However, there are two exceptions: (1) We do not have to introduce the $XT$ field
since the triplet Higgs field $H'_T$ can have mass around a few 
hundred GeV; (2) We do not have to introduce the vector-like particles
$(XL,~ \overline{XL})$ since the neutrino masses and mixings can arise
from the right-handed neutrino Majorana mass mixings. Then
both the tau lepton decays to electron/muon and the proton decays
to $\pi^0 e^+$ will be highly suppressed. 

The relevant renormalizable operators for the SM fermions are 
\begin{eqnarray}
-{\cal L} & = &
F_3 \overline{Xf} \Phi + N_k^c \overline{Xf} \Phi^{\dagger}+
\overline{XD} Q_i H + E_i^c XL H + \overline{Xf} XD U_T 
\nonumber\\ &&
+ \overline{f}_3 XD U_T
+ \overline{XD} Xf U_T^{\dagger} + D_i^c Xf U_T^{\dagger}
+ \overline{XL} \overline{Xf} U_D + \overline{XL}  \overline{f}_3 U_D
\nonumber\\ &&
+ Xf XL U_D^{\dagger} + Xf L_i U_D^{\dagger} 
+ \mu_{Xf3} \overline{f_3} Xf + \mu_{XDi} D_i^c XD + \mu_{XL_i} \overline{XL} L_i
\nonumber\\ &&
+ M_{Xf} \overline{Xf} Xf + M_{XD} \overline{XD} XD + M_{XL} \overline{XL} XL  + {\rm H.C.} 
~,~\,
\label{Complete-Operators}
\end{eqnarray}
where we neglect the Yukawa couplings for simplicity.
We assume that the mass terms  $M_{Xf}$, $M_{XD}$, and $M_{XL}$ are around 
1000~TeV, while the mass terms $\mu_{Xf3}$, $\mu_{XDi}$, and $\mu_{XL_i}$
are relatively small. This can be realized via rotations of the fields
since $Xf$, $XD$ and $\overline{XL}$  only couple to one linear
combinations of $\overline{Xf}/\overline{f}_3$, $\overline{XD}/D_i^c$,
$XL/L_i$, respectively. Because
the VEVs of $U_T$ and $U_D$ are around 1 TeV, the mixing terms from
$\overline{f}_3 XD U_T$, $D_i^c Xf U_T^{\dagger}$, 
$\overline{XL}  \overline{f}_3 U_D$, and $Xf L_i U_D^{\dagger}$ 
are small and negligible.

For the dimension-five operators in Eq.~(\ref{O-Dim-5}),
the $y^d_{i3}$ term can be generated from the above
renormalizable operators $D_i^c Xf U_T^{\dagger}$ and 
$F_3 \overline{Xf} \Phi$, the $y^e_{i3}$ term can be 
generated from the above renormalizable operators $E_i^c XL H$ and
$\overline{XL}  \overline{f}_3 U_D$, 
the $y^d_{3i}$ term can be generated from the above
renormalizable operators $\overline{f}_3 XD U_T$ and
 $\overline{XD} Q_i H$, and the $y^e_{3i}$ term can be 
generated from the above renormalizable operators $F_3 \overline{Xf} \Phi$ and
$Xf L_i U_D^{\dagger} $.

In addition, we can show that there are no up-type quark mixings
after we integrate out the vector-like particles. Let us explain the
point. The $SU(3)'_C\times SU(2)'_L \times U(1)'_Y$ gauge symmetry
can be formally embedded into a global $SU(5)'$ symmetry. Under
$SU(5)\times SU(5)'$, the bifundamental fields $U_T$ and $U_D$ form
$({\bf 5}, {\bf {\bar 5}})$ representation, the
vector-like particles $Xf$ and $\overline{Xf}$ respectively
form $({\bf 5}, {\bf { 1}})$ and $({\bf {\bar 5}}, {\bf 1})$ representations,
and the vector-like particles ($XD,~\overline{XL}$) and
($\overline{XD}, ~XL$)  respectively form $({\bf { 1}}, {\bf 5})$ and 
$({\bf { 1}}, {\bf {\bar 5}})$ representations. Because all these
fields are in the fundamental and/or anti-fundamental representations of $SU(5)$
and/or $SU(5)'$,  we cannot create the Yukawa interactions $10_f 10'_f 5_{H}$ or $10_f 10'_f 5_{H'}$ for the up-type quarks
after we integrate out the vector-like particles. 
Therefore, there is no proton decay problem.

\section{Phenomenology and signals at LHC}
In this section we discuss the production mechanism for the exotic gauge bosons in our model and focus on the $X_\mu$
and $Y_\mu$ vector bosons predicted in our model. These vector bosons carry both color and electroweak quantum numbers
and behave as leptoquarks as well as diquarks. As the gauge bosons have their origins in the gauge group $SU(5)$ which
unifies only the third generation, as far as its coupling to fermions is concerned, it couples only to the third generation quarks and 
leptons. However, it interacts with the gluon as well as to all the other electroweak gauge bosons of the SM which would help in 
producing these particles at collider experiments. As far as their production at hadron colliders is concerned the dominant contributions 
would come from the strongly interacting subprocesses and therefore one can neglect the sub-dominant contributions 
coming from electroweak gauge boson exchanges. Note that they will however be produced only through the exchange of electroweak 
gauge bosons at electron positron colliders such as the {\it International Linear Collider} (ILC) \cite{Djouadi:2007ik} or the CLIC 
\cite{Linssen:2012hp}, envisioned and proposed for the future. We restrict ourselves to the study of these gauge boson at the currently 
operational LHC at CERN and therefore only focus on the couplings of the $X_\mu$ and 
$Y_\mu$ vector bosons  with the gluons which would be relevant for its production at the LHC.  
The general form of the interaction 
can be derived from the Lagrangian given by \cite{blumlein:belyaev}
\begin{align}
{\mathcal L} = -\frac{1}{2}{\mathcal V}^{i\dagger}_{\mu\nu}{\mathcal V}_i^{\mu\nu} + M_V^2 V_\mu^{i\dagger} V^\mu_i
 -i g_s V_\mu^{i\dagger} {T^a_{ij}} V^j_\nu {\mathcal G}_a^{\mu\nu}
\label{eq:lagrangian}
\end{align} 
where $V\equiv X,Y$ and $T^a$ are the $SU(3)_c$ generators. The field strength tensors for the exotic vector fields $V_\mu$ and 
gluon $G_\mu^a$ are
\begin{align}
{\mathcal G}_a^{\mu\nu} &=\partial_\mu G_\nu^a - \partial_\nu G_\mu^a + g_s f^{abc} G_{\mu b} G_{\nu c}  \\ 
{\mathcal V}_i^{\mu\nu}  &= D_\mu^{ik} V_{\nu k} - D_\nu^{ik} V_{\mu k}
\end{align}
and the covariant derivative is defined as
\begin{align}
D_\mu^{ij} = \partial_\mu \delta^{ij} - i g_s T^{ij}_a G^a_\mu.
\end{align}
\begin{center}
\begin{figure}[!t]
\includegraphics[width=6.0in,height=1.5in]{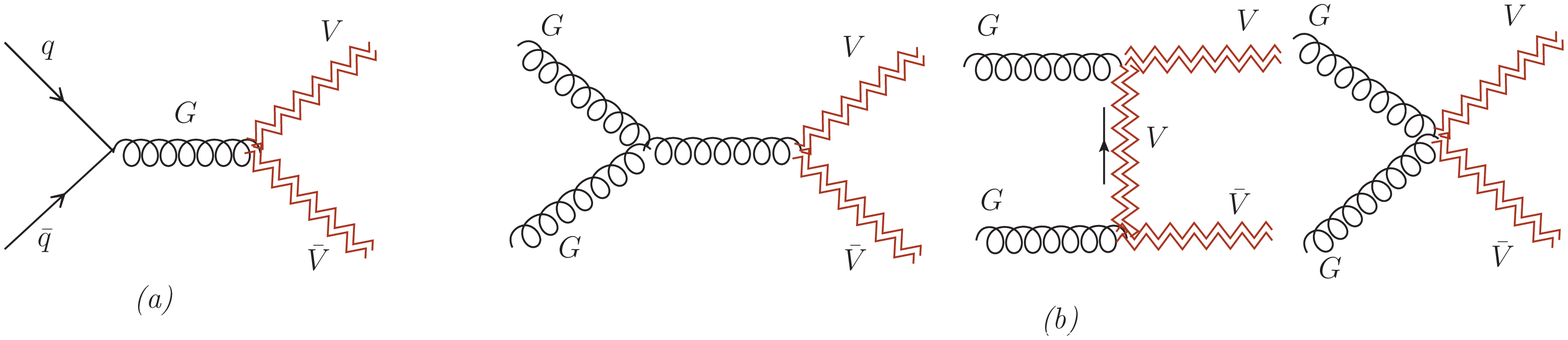}
\caption{The tree level Feynman diagrams which contribute to the pair production of the $X_\mu$ and $Y_\mu$ gauge bosons 
at the LHC, where both of them are denoted as $V_\mu$. The subprocesses that contribute are (a) $q\bar{q} \to V\bar{V}$ 
and (b) $G G \to  V\bar{V}$.  }
\label{fig:diagrams}
\end{figure}
\end{center}
Using the above Lagrangian we derive the Feynman rules for the interactions of the leptoquark gauge bosons $V\equiv X,Y$ with 
the gluon fields. These interactions then lead to the tree level Feynman diagrams as shown in 
Fig.(\ref{fig:diagrams}) which contribute to the pair production of these exotic particles at the LHC. 
\subsection{Calculation of cross sections}
\begin{figure}[t!]
\centering
\includegraphics[width=3.5in]{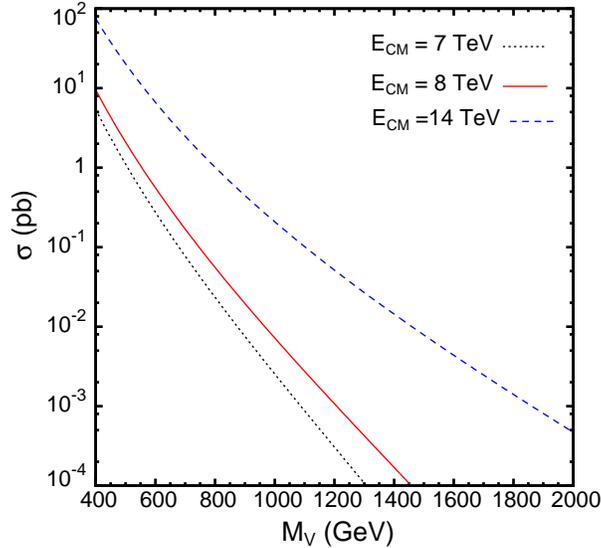}
\caption{\it The production cross sections for 
$p p \rightarrow V \bar{V}$ at the LHC as a function of 
leptoquark mass $M_V$ at center-of-mass energies, 
$E_{CM} = 7,8$ and $14$ TeV. We have chosen the scale as 
$Q=M_V$, the mass of the leptoquark.}
\label{prodfig}
\end{figure}

Using Feynman rules for the interaction vertices of the exotic gauge bosons with gluons derived 
from Eq.(\ref{eq:lagrangian}) we can write down the full spin and color averaged matrix amplitude square for the 
quark-antiquark annihilation subprocess  $q\bar{q} \to V \bar{V}$, (where $q\equiv u,d,c,s,b$ and $V\equiv X,Y$) as 
\begin{eqnarray*}
\overline{|{\mathcal M}|}^2_{q\bar{q}} = &&
\frac{g_s^4}{9 M_V^4 s^2} \left[-12 M_V^8-s^2 t (s+t)+4 M_V^6 (s+6 t)+2 M_V^2 s \left(2 s^2+3 s t+2 t^2\right) \right. \\ 
&& \left. -M_V^4 \left(17 s^2+20 s t+12 t^2\right)\right]
\end{eqnarray*}
while for the gluon induced  subprocess $GG \to V \bar{V}$, it is given by 
\begin{eqnarray*}
\overline{|{\mathcal M}|}^2_{GG} = && g_s^4
\left[\frac{9 M_V^4+4 s^2+9 s t+9 t^2-9 M_V^2 (s+2 t)}{24 s^2 \left(t-M_V^2\right)^2 \left(s+t-M_V^2\right)^2}\right]   \left[3 M_V^8+2 s^4-12 M_V^6 t+4 s^3 t  \right. \\
&& \left. +7 s^2 t^2+6 s t^3+3 t^4+M_V^4 \left(7 s^2+6 s t+18 t^2\right)-4 M_V^2 \left(s^3+2 s^2 t+3 s t^2+3 t^3\right)\right].
\end{eqnarray*}
Note that the Mandelstam variables $s$ and $t$ are defined in the parton frame of reference.  The
pair production cross section at the parton level is then easily obtained using the above 
expressions. To obtain the production cross section we convolute the parton level cross sections
$\hat{\sigma}(q_i\bar{q}_i\rightarrow V \bar{V})$ and $\hat{\sigma}(GG\rightarrow V \bar{V})$
 with the parton distribution functions (PDF).
\begin{equation}\label{prodcros}
\begin{split}
\sigma({pp \rightarrow V \bar{V}})=& 
            \left\{ \sum_{i=1}^{5}  \int dx_1 \int dx_2 
        ~\mathcal{F}_{q_i}(x_1,Q^2)\times \mathcal{F}_{\bar{q}_i}(x_2,Q^2) 
         \times \hat{\sigma}(q_i\bar{q}_i\rightarrow V \bar{V}) \right\} \\
  & + \int dx_1 \int dx_2 ~\mathcal{F}_g (x_1,Q^2) \times 
                           \mathcal{F}_g (x_2,Q^2) 
                  \times \hat{\sigma}(GG\rightarrow V \bar{V}), 
\end{split}
\end{equation}
where $\mathcal{F}_{q_i}$, $\mathcal{F}_{\bar{q}_i}$ and $\mathcal{F}_{g}$  
represent the respective PDF's for partons (quark, antiquark and gluons) in 
the colliding protons, while $Q$ is the factorization scale.  
In Fig.(\ref{prodfig}) we plot the leading-order production cross section 
for the process $p p \rightarrow V \bar{V}$  at center of mass energies 
of 7, 8 and 14 TeV  as a function of the leptoquark mass $M_V$.
We set the factorization scale $Q$ equal to $M_V$,  
and have used the {\tt CTEQ6L1} PDF  \cite{Pumplin:2002vw}. As seen from the plot, we find that 
the pair production cross section for both the $X$ and $Y$ leptoquark 
gauge bosons are quite big for significantly large values of their mass 
even at the 7 and 8 TeV runs of LHC. Thus one expects severe bounds on such
particle masses from experimental data. In an earlier work \cite{Chakdar:2012kd}, we 
had studied specific signals from the pair production of $X_\mu$ at LHC and put 
expected limits on its mass. This work was also followed up by the CMS experimental group
which placed comparable limits on such leptoquark vector bosons \cite{Chatrchyan:2012sv}
using collision data from the 7 TeV run of the LHC. We note that as both the $X$ and $Y$
leptoquark gauge bosons have identical masses, any limits on one of them invariably leads
to a similar limit on the other. Thus it is important to explore all possible signals that 
come from the pair productions of these particles. In this work we extend our earlier 
study by looking at the different signals from the pair productions of such particles 
at LHC with center-of-mass energies of 8 TeV and 14 TeV. We note that at the 14 TeV run of LHC 
the production cross section for the leptoquark gauge bosons is significantly enhanced and 
would therefore improve the reach for such particle searches.
\subsection{Calculation of decays of the $X_\mu$ and $Y_\mu$ gauge bosons}
\begin{figure}[ht!]
\centering
\includegraphics[width=2.1in]{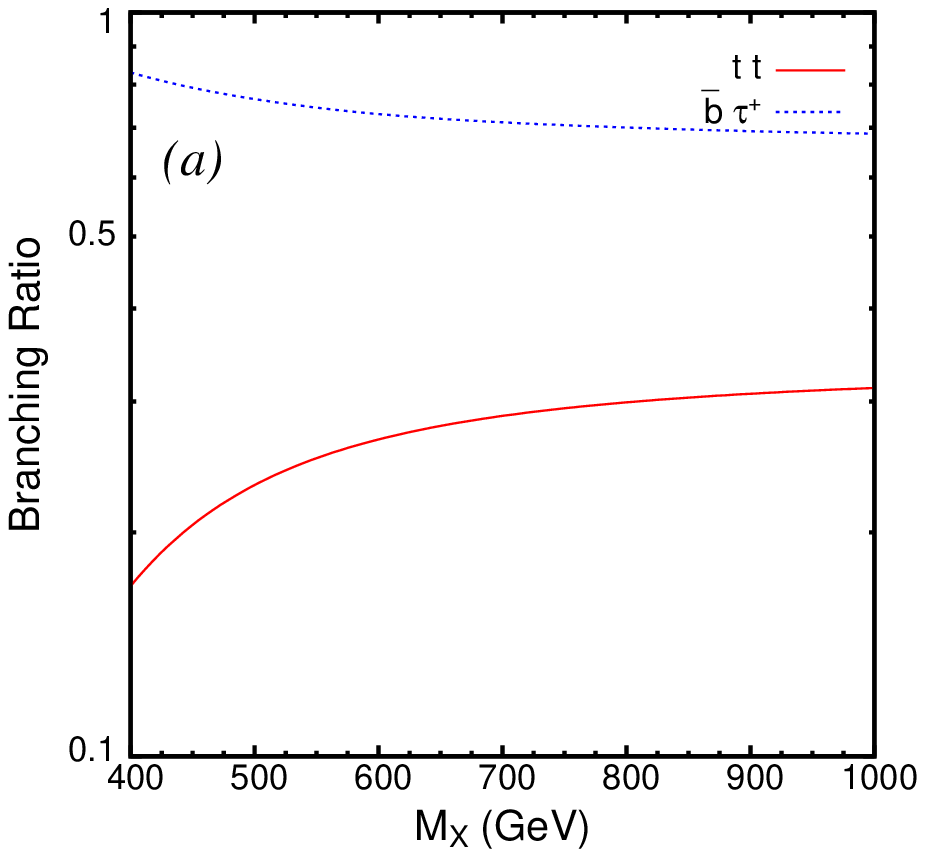}
\includegraphics[width=2.1in]{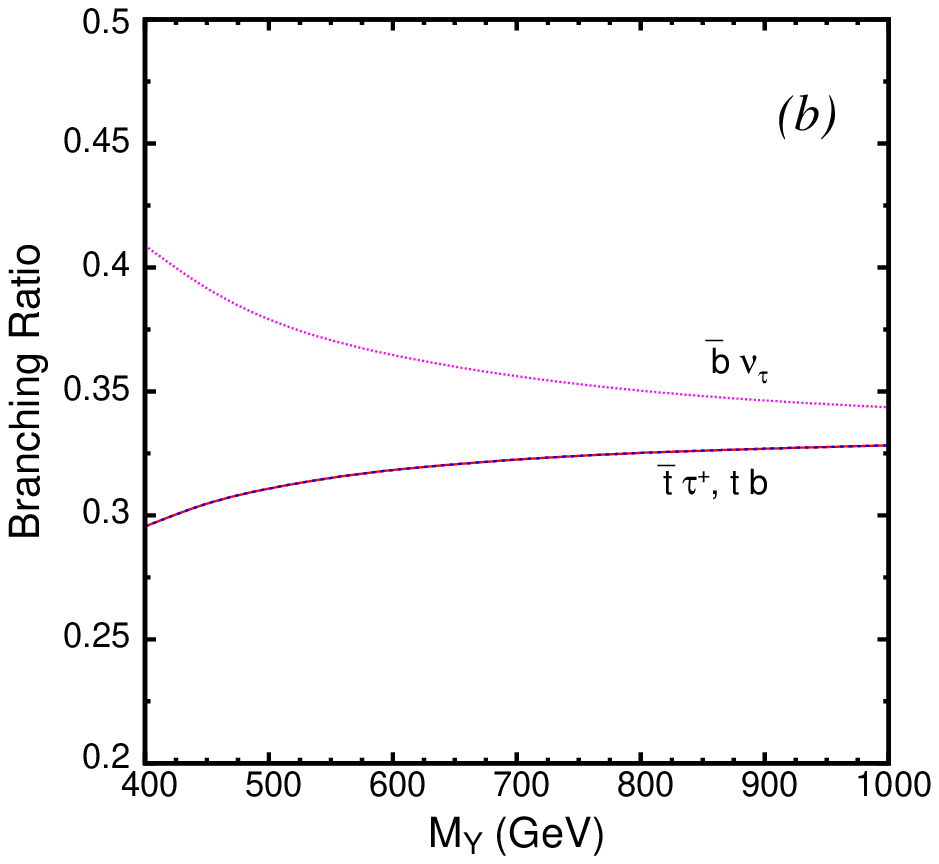}
\includegraphics[width=2.0in]{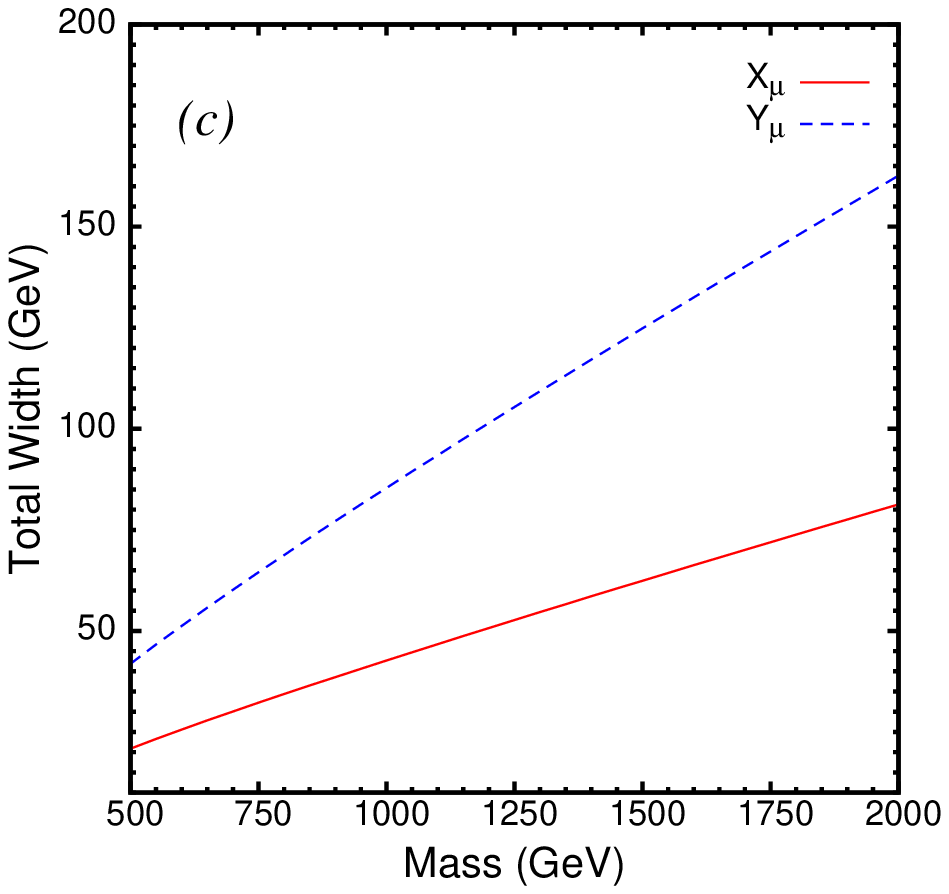}
\caption{\it Illustrating the decay branching fractions of the leptoquark gauge bosons (a) $X_\mu$, 
(b) $Y_\mu$ and, (c) the total widths as a function of their mass.}
\label{decayfig}
\end{figure}

To study the possible signals for the leptoquark gauge bosons, we need to know their decay 
properties. Since the third family of fermions is only charged under the gauge group $SU(5)$,
these leptoquark gauge bosons which come from the $SU(5)$ gauge fields are only coupled
to the third generation fermion fields. The interaction Lagrangian of the leptoquark gauge bosons
$X_\mu$ and $Y_\mu$ with the third generation fermions is given by \cite{Langacker:1980js},
\begin{align}
\mathcal{L}_G = & \frac{g_5}{\sqrt{2}} \bar{X}_\mu^\alpha \big[ \bar{b}_{R\alpha}\gamma^\mu \tau_R^+ +
\bar{b}_{L\alpha}\gamma^\mu \tau_L^+ + \epsilon^{\beta\gamma}_{\alpha} \bar{t}^c_{L\gamma} 
\gamma^\mu t_{L\beta} \big]  \nonumber \\ 
+ & \frac{g_5}{\sqrt{2}} \bar{Y}_\mu^\alpha \big[- \bar{b}_{R\alpha}\gamma^\mu \nu_R^c -
\bar{t}_{L\alpha}\gamma^\mu \tau_L^+ + \epsilon^{\beta\gamma}_{\alpha} \bar{t}^c_{L\gamma} 
\gamma^\mu b_{L\beta} \big] + H.C.
\label{eq:decaylag}
\end{align}
Using the above interaction Lagrangian, we can calculate the explicit decay modes of the leptoquark gauge bosons,
where $X_\mu$ decays to a top quark pair ($tt$) or anti-bottom quark + positively charged tau lepton ($\bar{b}\tau^+$) while $Y_\mu$ has three decay modes to anti-bottom quark + a tau-neutrino ($\bar{b}\nu_\tau$), anti-top quark + positively charged tau ($\bar{t}\tau^+$) or top quark + bottom quark ($tb$).
The partial decay width for each mode calculated using Eq.(\ref{eq:decaylag}) is then given by
\begin{align}
 \Gamma (X \to t t~~~) &= \frac{g_5^2 M_X}{24\pi} \left(1-\frac{m_t^2}{M_X^2} \right) 
                                          \left(1 -\frac{4m_t^2}{M_X^2} \right)^{1/2} \nonumber \\
 \Gamma (X \rightarrow \bar{b} \tau^+) &= \frac{g_5^2 M_X}{12\pi}   \\ \nonumber \\
 \Gamma (Y \to \bar{t} \tau^+) &= \frac{g_5^2 M_Y}{24\pi} \left(1-\frac{m_t^2}{M_Y^2}\right)^2 
                                                       \left(2 +\frac{m_t^2}{M_Y^2} \right) \nonumber \\
 \Gamma (Y \to \bar{b} \nu_\tau) &=\frac{g_5^2 M_Y}{12\pi}  \nonumber \\
 \Gamma (Y \to t b~) &=  \frac{g_5^2 M_Y}{24\pi} \left(1-\frac{m_t^2}{M_Y^2} \right)^2 
                                                      \left (2 +\frac{m_t^2}{M_Y^2} \right)
\end{align}
where $g_5$ is the $SU(5)$ gauge coupling and we have only kept the top quark mass ($m_t$) and neglected the other fermion masses.
We plot the branching fractions of the leptoquark gauge bosons decays as well as their total widths, as 
shown in Fig.(\ref{decayfig}). It is interesting to note that while the 
$X_\mu$ decays dominantly to $\bar{b}\tau^+$, it also has a substantial branching fraction to a pair of 
same sign top quarks. For very large values of the mass $M_X$ of the $X_\mu$, when the mass of the top
quark can be neglected, we find that $\frac{\Gamma(X\to \bar{b}\tau^+)}{\Gamma(X\to tt)} \simeq 2$. 
For the $Y_\mu$ leptoquark gauge boson we find that for smaller values of its mass it has the dominant 
decay fraction to $\bar{b}\nu_\tau$ while its decay to $\bar{t}\tau^+$ and $tb$ are equal. But for $M_Y$
quite large such that the top quark mass may be neglected, all $Y_\mu$ decay modes have the same branching
probability of 1/3.
With the knowledge of the decay modes of the leptoquark gauge bosons and the branching fractions for the
decays we can now analyze all the different final states that we expect from the pair production of these 
leptoquarks at the LHC. Note that the total decay widths ($\Gamma_V$) for the  
leptoquark gauge bosons are such that  $\Gamma_V < 0.1 M_V$  and we have therefore used the
narrow width approximation (NWA) which proves to be a useful tool in simplifying the analysis without 
introducing large errors \cite{Berdine:2007uv}. We have fixed the $g_5$ coupling to the value of the strong
coupling constant ($g_s$) throughout the analysis.

\subsection{Signals at LHC}
In Ref.\cite{Chakdar:2012kd} we studied the signals for the pair production of the $X_\mu$ leptoquark gauge bosons
and their subsequent decays into the dominant mode $\bar{b}\tau^+$ at LHC center of mass energies of 7 TeV and 8 TeV. 
The final state signal was $b\bar{b}\tau^+\tau^-$ with all the four particles being detected in the respective flavor tagged
mode. It was observed that the signal stands out as resonances in the invariant mass distribution of the $\tau$ lepton paired 
with the $b$ jets against the continuum SM background, provided all the four final state particles carried significant transverse 
momenta. Using this signal a phenomenological prediction on the LHC reach was made on the mass of the $X_\mu$ which 
has subsequently been estimated as 760 GeV at 95 \% C.L. by the CMS Collaboration \cite{Chatrchyan:2012sv} at LHC with 
7 TeV center of mass energy. As our model predicts another decay mode (to top quark pairs) for the $X_\mu$ gauge boson, 
where the $X_\mu$ behaves as a {\it diquark}, carrying quantum numbers of two quarks, it is of extreme importance to 
be able to highlight this characteristic which distinguishes this particle from the usual leptoquark particles. Establishing the existence of both decay modes  is needed to show that these interactions are both baryon and lepton number violating. It is also worth 
pointing out that a similarly massive $Y_\mu$ in the spectrum which couples as strongly to the gluons as the $X_\mu$ will
also be produced with similar rates and needs to be studied in tandem with the production of the $X_\mu$ particles at the 
LHC.

We now consider all decay modes of both the $X_\mu$ and $Y_\mu$ and discuss final states which is then studied against the 
SM backgrounds. For the pair production process of $X_\mu$ gauge bosons,
where $X \to \bar{b}\tau^+,tt$ we have the following different final states given as
\begin{align}
p p \longrightarrow X \bar{X} \longrightarrow  \bar{b}\tau^+ b \tau^- ,~ t t b \tau^- ,~ \bar{b}\tau^+ \bar{t} \bar{t},
                                               ~  t t \bar{t} \bar{t}.  
\label{eq:xdecay}
\end{align}
The top quark would further decay, either semileptonically or hadronically to give multi-lepton and high jet multiplicity final states.
For our purposes, if we assume that the top quarks could be reconstructed with some reasonable efficiency in either modes, we can
just focus on the above mentioned final state signal.  Similarly for the pair production of the  $Y_\mu$ gauge bosons,
where $Y \to \bar{b} \nu_\tau, \bar{t}\tau^+, t b$ we get the following set of final states given as
\begin{align}
p p \longrightarrow Y \bar{Y} \longrightarrow & ~ b \bar{b} \slashed{E}_T, ~\bar{b} t \tau^- \slashed{E}_T, ~\bar{t} b \tau^+ \slashed{E}_T,
                                                                              ~\bar{b}\bar{b}\bar{t}\slashed{E}_T,~ b b t \slashed{E}_T,  \nn \\
                                              \hookrightarrow  & ~ t \bar{t}\tau^+\tau^-,~\bar{t}\bar{t}\bar{b}\tau^+,~ t t b \tau^-, ~t \bar{t} b \bar{b}.
\label{eq:ydecay}
\end{align}
Note that both the $X_\mu$ and $Y_\mu$ gauge boson productions at the LHC leads to a rich range of diverse final states which lead to 
many multi-particle signals and would lead to distinct resonances in the invariant mass distributions in some pairs corresponding to 
the mass of the $X_\mu$ and $Y_\mu$ states. Notably we find that each particular event rate is fixed once the model parameters have been
fixed, which in our case is the mass of the leptoquark gauge bosons while its coupling strength to the gluons 
has been fixed to be the 
strong coupling constant. Thus the success of the model is not dependent on an observation in only one 
particular final state but that observation needs to be complemented simultaneously in various other channels as listed above in Eqs.(\ref{eq:xdecay}) and (\ref{eq:ydecay}). Thus the study on all simultaneous channels 
deserves merit as it will be able to confirm or falsify the model in question.   

We now consider different final states and analyze the signals against the SM background. As we expect that the new gauge 
bosons when produced on-shell will decay to specific final state products, this would lead to a bump in the invariant mass distribution of 
the decay products. Keeping this in mind, it is instructive to first consider the most likely signals where the resonances would be observable.
Based on the decay channels and final states listed in Eqs.(\ref{eq:xdecay})--(\ref{eq:ydecay}) one 
should consider the ($b\tau$) mode for the
$X_\mu$ gauge bosons while the ($tb$ and $t\tau$) mode looks the more promising for the $Y_\mu$ resonance searches. The other modes either involve neutrinos or more than a single top
quark in the final state, which further decays either semileptonically or hadronically and therefore makes 
it more tasking to reconstruct the leptoquark gauge boson mass. 
However, we must emphasize that for measuring the electric charge of these gauge bosons one definitely requires that the $X_\mu$ resonance is observed in the invariant mass distribution of 
same-sign top quark pair ($tt$) while the $Y_\mu$ resonance is observed in the ($t\tau^-$) 
final state or its charge conjugate mode. Notwithstanding the fact that reconstructing the $tt$
state would be challenging, it would definitely lead to a very interesting observation. Final states
involving $b$ jets require measuring the $b$ jet charge which looks to be more difficult and 
hence not a desired mode to get information on the charge of the exotic gauge bosons. 

\begin{table}[!h]
\begin{center}
\begin{tabular}{|c|c|c|c|}
\hline ${\mathcal Signal}$& ${\mathcal SM}$ & ${\mathcal Signal}$& ${\mathcal SM}$  \\ 
\hline  $2b\tau^+\tau^-$ & $2b\tau^+\tau^-; ~ 2j\tau^+\tau^-;~ b j\tau^+\tau^-$  
       &  $ttb\tau^-,~\bar{t}\bar{t}b\tau^+$ & -- \\ 
\hline  $tt\bar{t}\bar{t}$ & $tt\bar{t}\bar{t}$ 
       &  $t\bar{t}\tau^+\tau^-$ & $t\bar{t}\tau^+\tau^-$ \\
\hline  $2b t\bar{t}$ & $2b t\bar{t}; ~ 2j t\bar{t}; ~ b j t\bar{t}$ 
       &  $2b\slashed{E}_T$ & $2b\slashed{E}_T; ~ 2j \slashed{E}_T; ~ j b \slashed{E}_T$ \\
\hline $b t \tau^- \slashed{E}_T$ & $b t \tau^- \slashed{E}_T; ~ j t \tau^- \slashed{E}_T$ 
       & $b \bar{t} \tau^+\slashed{E}_T$ & $b \bar{t} \tau^+\slashed{E}_T;~ j\bar{t} \tau^+\slashed{E}_T$ \\
\hline  $2b t \slashed{E}_T$ & $ 2j t\slashed{E}_T; ~ b j t\slashed{E}_T$   
       &  $2b\bar{t}\slashed{E}_T$ & $2j\bar{t}\slashed{E}_T; ~bj\bar{t}\slashed{E}_T$ \\
\hline
\end{tabular}
\caption{\textit{Illustrating the final state signals and the corresponding SM background 
subprocesses. Note that $\slashed{E}_T$ for the SM subprocesses represents one or more neutrinos 
in the final state.}} \label{tab:sigbkg}
\end{center}
\end{table}
In Table \ref{tab:sigbkg} we list the relevant SM background subprocesses  that 
we have considered for each set of final states for the signal. Note that we do not make a distinction
between the $b$ and $\bar{b}$ but we distinguish between a $\tau^+$ and $\tau^-$ by assuming exact 
charge measurement will be possible. We also distinguish between a top quark and anti-top quark assuming
that they will be reconstructed with their respective charge identifications from its semileptonic decay modes.
We associate an efficiency factor of $\varepsilon_t$ with this reconstruction. For final state signals not involving neutrinos we have not considered SM subprocesses with $\slashed{E}_T$ as they will involve  
extra electroweak vertices which suppress the contributions and further requirements on missing 
transverse momenta would make these contributions too small to take into further consideration.  
We highlight the above mentioned invariant mass distributions in our model for a few choices 
of the $X_\mu$ and $Y_\mu$ gauge boson masses considered at two different center of mass energies for the LHC. We focus our attention to the recently concluded 8 TeV run and the
proposed upgrade in energy of 14 TeV for the LHC. As a current limit of 760 GeV exists on the
leptoquark gauge boson mass from the CMS analysis \cite{Chatrchyan:2012sv} we choose 
a mass of 800 GeV to show the distributions at the 8 TeV run of LHC while a larger mass of 
1 TeV is chosen to highlight the signal distributions at the 14 TeV run. We note that there are 
more than one set of final states where a particular resonance could be observed in the invariant
mass distributions and so we consider the scenario where we look at a few definite invariant mass
distributions in individual final state modes listed in Eqs.(\ref{eq:xdecay}) and (\ref{eq:ydecay}). 
We list below the pair of final
state particles for which the invariant mass distribution is considered, motivated by favored modes for 
reconstructing the mass and the charge of the $X_\mu$ and $Y_\mu$ gauge bosons.

\begin{itemize}
\item [{\bf (C1)}] Invariant mass distribution of $b\tau^-$ coming from the final states $\bar{b}\tau^+ b \tau^- ,~ t t b \tau^-$. This is the most favorable mode for reconstructing the $X_\mu$ resonance.
\item [{\bf (C2)}] Invariant mass distribution of same sign top quark pair $tt$ coming from the final states
$ttb\tau^-,~tt\bar{t}\bar{t}$. The reconstruction of the leptoquark mass in this mode, although difficult, is
essential in measuring the charge of the $X_\mu$. 
\item [{\bf (C3)}] Invariant mass distribution of $tb$ coming from the final states $t\bar{t}b\bar{b},~ttb\tau^-,~t\bar{b}\tau^-\slashed{E}_T,~bbt\slashed{E}_T$. This is one of the 
favorable modes for reconstructing the $Y_\mu$ resonance.
\item [{\bf (C4)}] Invariant mass distribution of $t\tau^-$ coming from the final states $t\bar{t}\tau^-\tau^+,~tt b\tau^-,~\bar{b}t\tau^-\slashed{E}_T$. This mode is essential to measure the charge of the $Y_\mu$. 
Note that the $t\tau^-$ resonance corresponds to the charge conjugate mode of $Y_\mu$.   
\end{itemize}

We shall now discuss the signal and the associated SM backgrounds for the 
list of resonances given by {\bf C1--C4}. Note that the signal subprocesses which contribute 
to give a $b\tau^-$ final state as listed in {\bf (C1)} come from both $X_\mu$ and $Y_\mu$
pair productions. However the resonant distribution only happens for the $X_\mu$ production
while the $Y_\mu$ contribution acts to smear out the resonance although it does contribute 
in enhancing the signal over the SM background. A further smearing effect would come if the 
$t\bar{b}\tau^-\slashed{E}_T$ signal is included. But we can reject that contribution by demanding 
that we don't include events with large missing transverse momenta in the final state when 
reconstructing the $b\tau^-$ invariant mass. As discussed in Ref.\cite{Chakdar:2012kd} 
the dominant background for the resonant signal in the $b\tau$ channel comes from 
$pp\to 2b2\tau,4b,2j2b,2j2\tau,4j,t\bar{t}$ where $j=u,d,s,c$ when we consider the 
signal coming from the pair production of $X_\mu$ which then decay in the $b\tau$ mode 
to give a $2b\tau^+\tau^-$ final state.  The light jet final 
states in the SM can be mistaged as $\tau$ or $b$ jets and thus form a significant source for 
the background due to the large cross sections at LHC, as they are dominantly produced through 
strong interactions. Guided by previous analysis \cite{Chakdar:2012kd}, we 
note that a very strong requirement on the transverse momenta for the $b$ jet and the $\tau$
lepton is very helpful in suppressing the SM background. The SM background has been 
estimated using {\tt Madgraph 5} \cite{mad5}. In this analysis we further restrict the number of 
SM background sub-processes that contribute to the final state with $b\tau^-$ by 
demanding that the tau charge is measured. Therefore we neglect the 
contributions coming from jets that fake a tau. For example, when we consider 
the final state as $2b\tau^+\tau^-$ and demand that the tau lepton is tagged as 
well as its charge measured, we include $pp\to 2b2\tau,2j2\tau,t\bar{t}$ as the 
dominant SM processes for the background.

We have used two values for the leptoquark gauge boson ($V\equiv X,Y$) masses, $M_V=800$ 
GeV at LHC with center of mass energy 8 TeV and $M_V=1$ TeV at LHC with center of mass 
energy 14 TeV to highlight the signal cross sections and differential distributions for 
invariant mass. We set the factorization and renormalization scale ($Q=M_Z$) to the mass of the $Z$ boson and
also use the strong coupling constant value of $\alpha_s$ evaluated at the $Z$ boson mass.  
Note that we have evaluated the individual signals as listed in Eqs.(\ref{eq:xdecay}) and (\ref{eq:ydecay}) 
against their specific backgrounds independently. 
We have assumed in our analysis that the top quark and the anti-top quark are reconstructed
with good efficiencies which we can parameterize as $\varepsilon_t$. Note that we have used the 
following efficiencies for $b$ and $\tau$ tagging, $\epsilon_b=\epsilon_\tau=0.5$ while we assume a mistag rate 
for light jets to be tagged as $b$ jets as 1\% and $c$ jets tagged as $b$ jets to be 10\%. All our results here
are done at the parton level and therefore to account for the detector resolutions for energy measurement 
of particles, we have used a Gaussian smearing of the jet and $\tau$ energies with an energy resolution given 
by $\Delta E/E = 0.8/\sqrt{E~(GeV)}$ and $\Delta E/E = 0.15/\sqrt{E~(GeV)}$ respectively when analyzing 
the signal events. 

In Table \ref{tab:cuts} we list the kinematic selection cuts on the events. As the primary decay modes 
of the heavy leptoquark gauge bosons will have very large transverse momenta we put strong cuts on them. 
This helps in suppressing the SM background while it does not have any significant effect on the signal 
events. The cuts on $\slashed{E}_T$ applies only to final
states with neutrinos in the decay chain while the $\Delta R_{ij}$ cut is on any pair of visible particles. 
The invariant mass cut $M_{jj}$ is on any pair of jets in the final state.
\begin{table}[!t]
\begin{center}
\begin{tabular}{|c|c|c|}
\hline {\bf Variable} &Cut ${\mathcal C}_1$ at 8 TeV & Cut ${\mathcal C}_2$ at 14 TeV \\ 
\hline $p_T^{\tau,b,j}$ & $>80$ GeV & $>200$ GeV \\ 
\hline $\slashed{E}_T$ & $>100$ GeV & $>200$ GeV \\ 
\hline $|\eta|$ & $<2.5$  & $<2.5$ \\ 
\hline $\Delta R_{ij}$ & $>0.4$ & $>0.4$ \\
\hline $M_{jj,\tau^+\tau^-}$ & $>5$ GeV & $>5$ GeV \\
\hline
\end{tabular}
\caption{\it Two different set of cuts, ${\mathcal C}_1$ at LHC with $\sqrt{s}=8$ TeV and 
${\mathcal C}_2$ at LHC with $\sqrt{s}=14$ TeV, imposed on the final states listed in 
Eqs.(\ref{eq:xdecay})--(\ref{eq:ydecay}) where the cuts on $\slashed{E}_T$ applies only to final
states with neutrinos in the decay chain.} \label{tab:cuts}
\end{center}
\end{table}

\begin{figure}[ht!]
\centering
\includegraphics[height=2.2in]{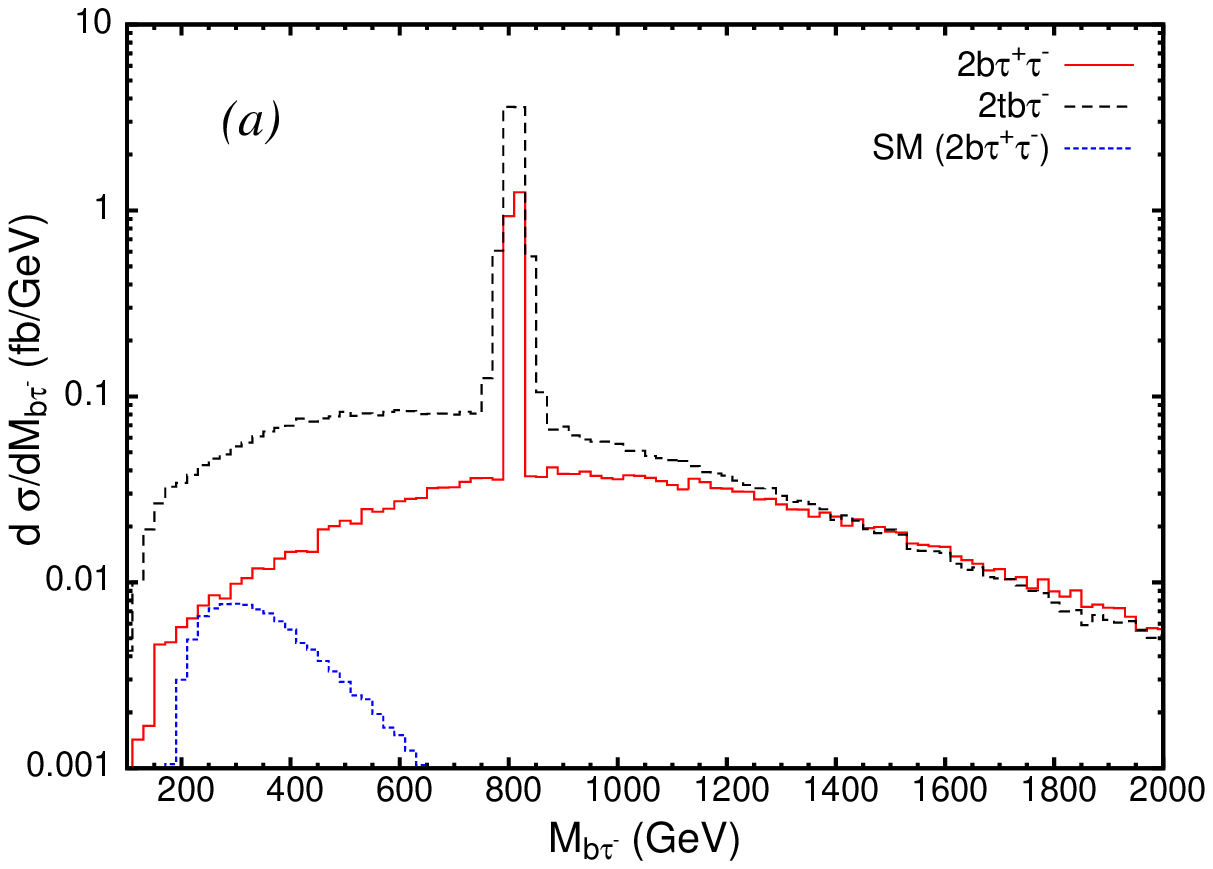}
\includegraphics[height=2.2in]{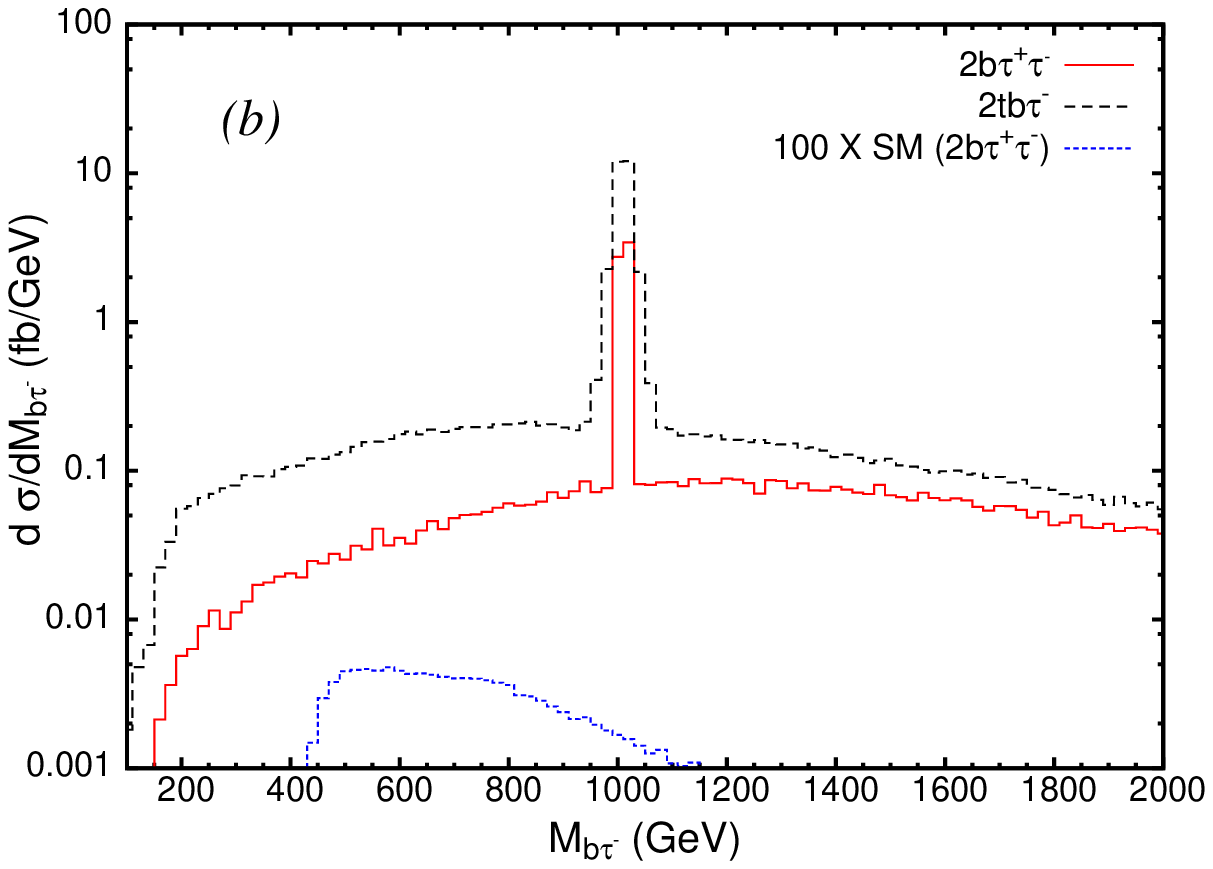}
\caption{\it Invariant mass distribution of $b\tau^-$ for the signal and SM background for two 
different choices of leptoquark gauge boson mass, (a) $M_V = 800$ GeV considered at LHC with 
$\sqrt{s}=8$ TeV and (b) $M_V = 1$ TeV considered at LHC with $\sqrt{s}=14$ TeV.}
\label{fig:btau}
\end{figure}

With the above set of kinematic selection on the final state events we evaluate the signal
cross sections and the corresponding SM background given in Table \ref{tab:sigbkg}. We 
first consider the resonance given by {\bf (C1)} and show the invariant mass distribution of $b\tau^-$ 
in Fig. \ref{fig:btau}. We must point out here that the $\tau^-$ is paired with the $b$ jet which has the 
leading transverse momenta in case there exist more than one tagged $b$ jets. 
After including the efficiency factors $\epsilon_b$ and $\epsilon_\tau$ associated with tagging the 
$b$ and $\tau$ jets and mistag rates, we estimate the signal cross section in the 
$2b\tau^+\tau^-$ mode as $4.23 ~fb$ for $M_{X,Y}=800$ GeV at LHC with $\sqrt{s}=8$ TeV and 
$12.05~fb$ 
for $M_{X,Y}=1$ TeV at LHC with $\sqrt{s}=14$ TeV. In Fig.\ref{fig:btau} we plot the invariant mass 
distribution for the signal. The dominant SM background are given by the following subprocesses, 
$\sigma (2b\tau^+\tau^-) \simeq 1.8 ~fb, ~\sigma (2c\tau^+\tau^-) \simeq 1.6 ~fb$ and 
$\sigma (2j\tau^+\tau^-) \simeq 167.6 ~fb$ which after including the efficiency factors, mistag rates 
is added to give $0.119~fb$. This is plotted in Fig.(\ref{fig:btau}) as ``SM $(2b\tau^+\tau^-)$". The 
corresponding SM background at 14 TeV center of mass energy is much more suppressed 
($\sim 0.002 ~fb$) because of the strong requirement on the transverse momenta of the jets and the 
charged tau leptons. The signal is clearly seen to stand out as resonance and one therefore expects 
this particular mode to be very  favorable in searching for the $X_\mu$ resonance by suppressing 
the SM background by demanding $\tau$ lepton charge identification which gets rid of the large all 
jet background.  Another mode for the $b\tau^-$ resonance which has completely negligible SM 
background, is for the final state $ttb\tau^-$. There are two different sources for the signal in this 
case, one which corresponds to the final states coming from the $X\bar{X}$ pair production while 
the other from the $Y\bar{Y}$ pair production. As the $Y\bar{Y}$ contribution does not lead to a 
resonance in the $b\tau^-$ mode, it will act to smear out the resonance as compared to that seen 
for the $2b\tau^+\tau^-$ final state. This is evident in Fig.(\ref{fig:btau}) where the width of the 
resonance is seen to spread out in more invariant mass bins for the $ttb\tau^-$  final state.  
Assuming a top reconstruction with an efficiency of $\varepsilon_t$ we find that the signal cross 
section from $X\bar{X}$ for $M_X=800~(1000)$ GeV at LHC with $\sqrt{s}=8~(14)$ TeV is 
$8.19 ~(28.23) \times \varepsilon_t^2~fb$ while the signal cross section from $Y\bar{Y}$ for 
$M_Y=800~(1000)$ GeV at LHC with $\sqrt{s}=8~(14)$ TeV is 
$4.04 ~(13.8) \times\varepsilon_t^2~fb$. Note that the $\tau$ and $b$ tagging 
efficiencies have been already included.
In Fig.(\ref{fig:btau}) we have assumed $\varepsilon_t=1$ for illustration purposes. Therefore the 
efficacy of the signal with the same sign top pairs in the final state is dependent on the inherent purity 
of the top reconstruction at experiments.

\begin{figure}[ht!]
\centering
\includegraphics[height=2.2in]{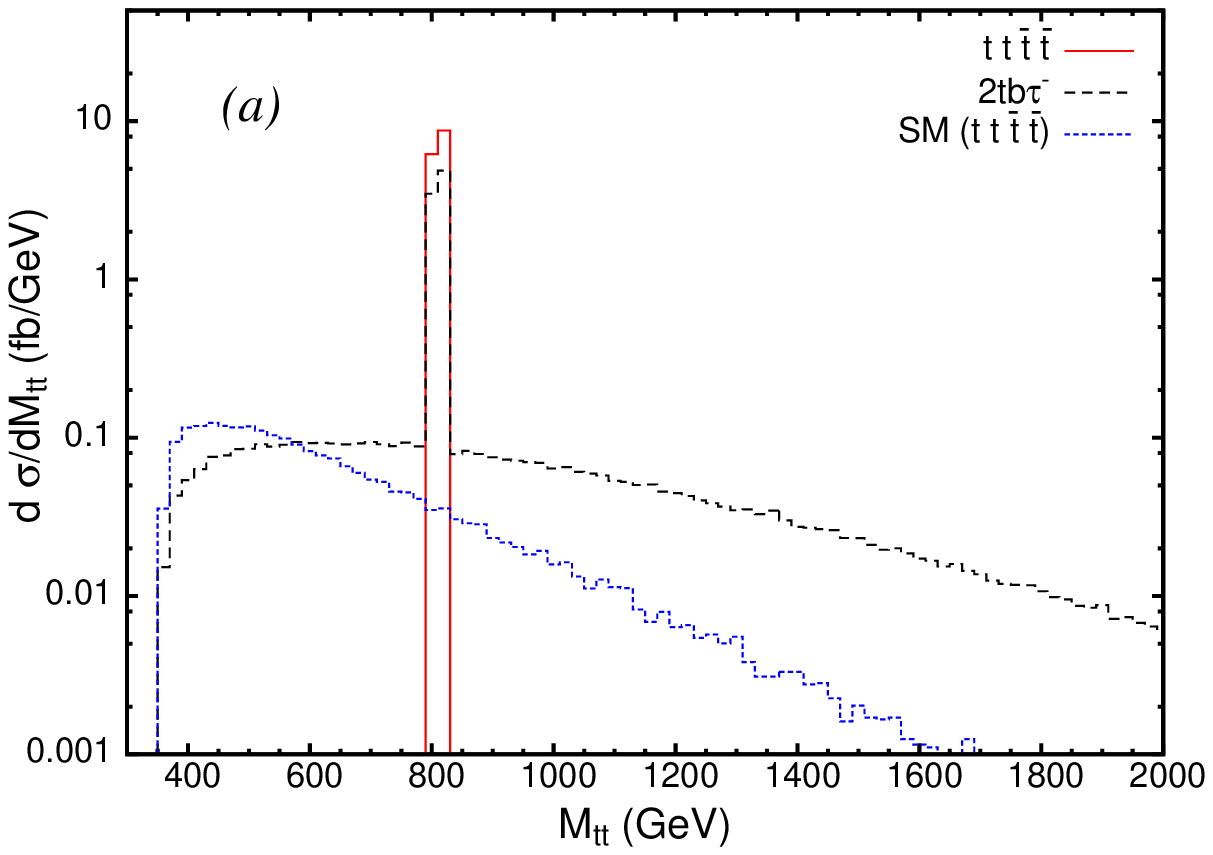}
\includegraphics[height=2.2in]{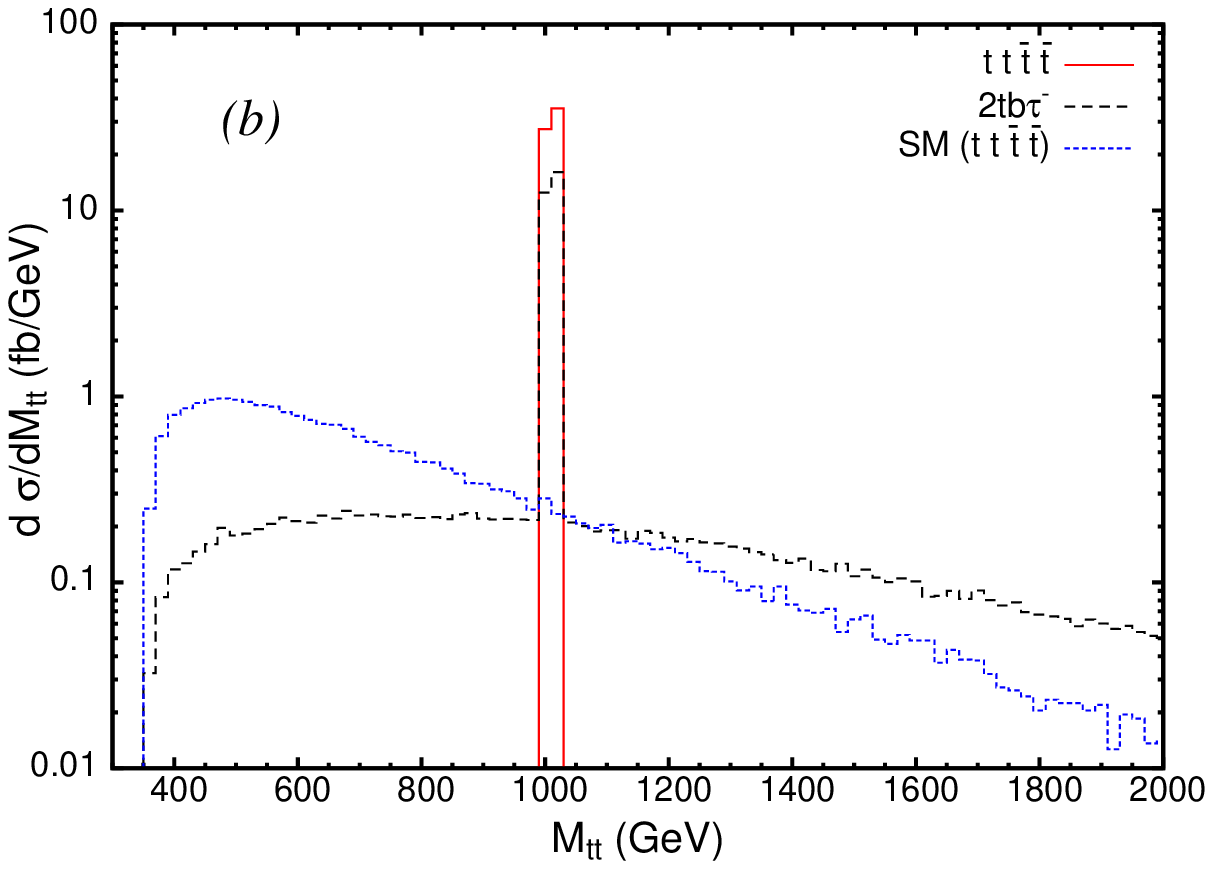}
\caption{\it Invariant mass distribution of same sign top pair $tt$ for the signal and SM background 
for two different choices of leptoquark gauge boson mass, (a) $M_V = 800$ GeV considered at LHC 
with $\sqrt{s}=8$ TeV and (b) $M_V = 1$ TeV considered at LHC with $\sqrt{s}=14$ TeV.}
\label{fig:tt}
\end{figure}

We now consider the resonance given by {\bf (C2)} and show the invariant mass distribution of 
the same sign top pair $tt$ in Fig. (\ref{fig:tt}). As pointed out earlier, this 
mode is necessary to measure the charge of the $X_\mu$ leptoquark gauge boson 
mass. A resonant bump in the same sign top pair invariant mass distribution would be a clear 
indication of a particle decaying into two same sign top quarks and therefore 
give a strong indication that the particle carries 4/3 electric charge and has quantum numbers of a
diquark.  
The signal is again considered for two different set of final states, both of which show an invariant 
mass peak in the same sign top quark pair. In the $t\bar{t}t\bar{t}$ final state the signal cross section
comes solely from the pair production of the $X\bar{X}$ gauge bosons. As we have assumed a 
reconstruction efficiency for the top quarks as $\varepsilon_t$, the cross section for 
$M_X=800~(1000)$ GeV at LHC with $\sqrt{s}=8~(14)$ TeV is $14.93~(62.81) \times 
\varepsilon_t^4 ~fb$. The SM background for the same subprocess is 
$2.31~(24.34) \times \varepsilon_t^4 ~fb$ at LHC with $\sqrt{s}=8~(14)$ TeV. 
Although the strength of the signal crucially depends on the reconstruction 
efficiency, even a low efficiency in the long run will lead to a very important 
observation provided similar resonances are observed in the $b\tau^-$  or $b\tau^+$ final 
states. The other final state which shows a bump in $tt$ invariant mass is $ttb\tau^-$  
and its strength was already discussed for Fig.(\ref{fig:btau}). Note that again 
the $Y\bar{Y}$ contribution does not help the resonance, but is effective in 
enhancing the signal in this mode.

\begin{figure}[ht!]
\centering
\includegraphics[height=2.2in]{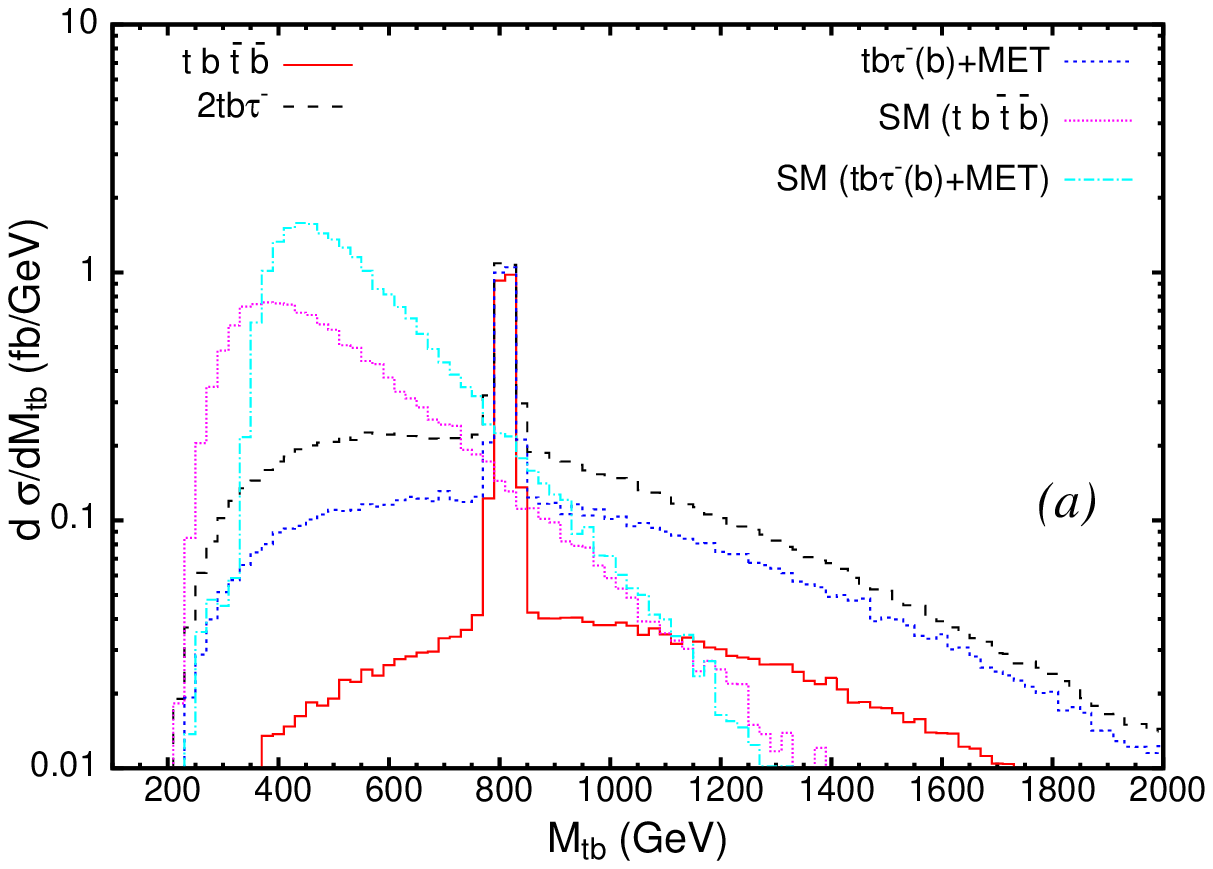}
\includegraphics[height=2.2in]{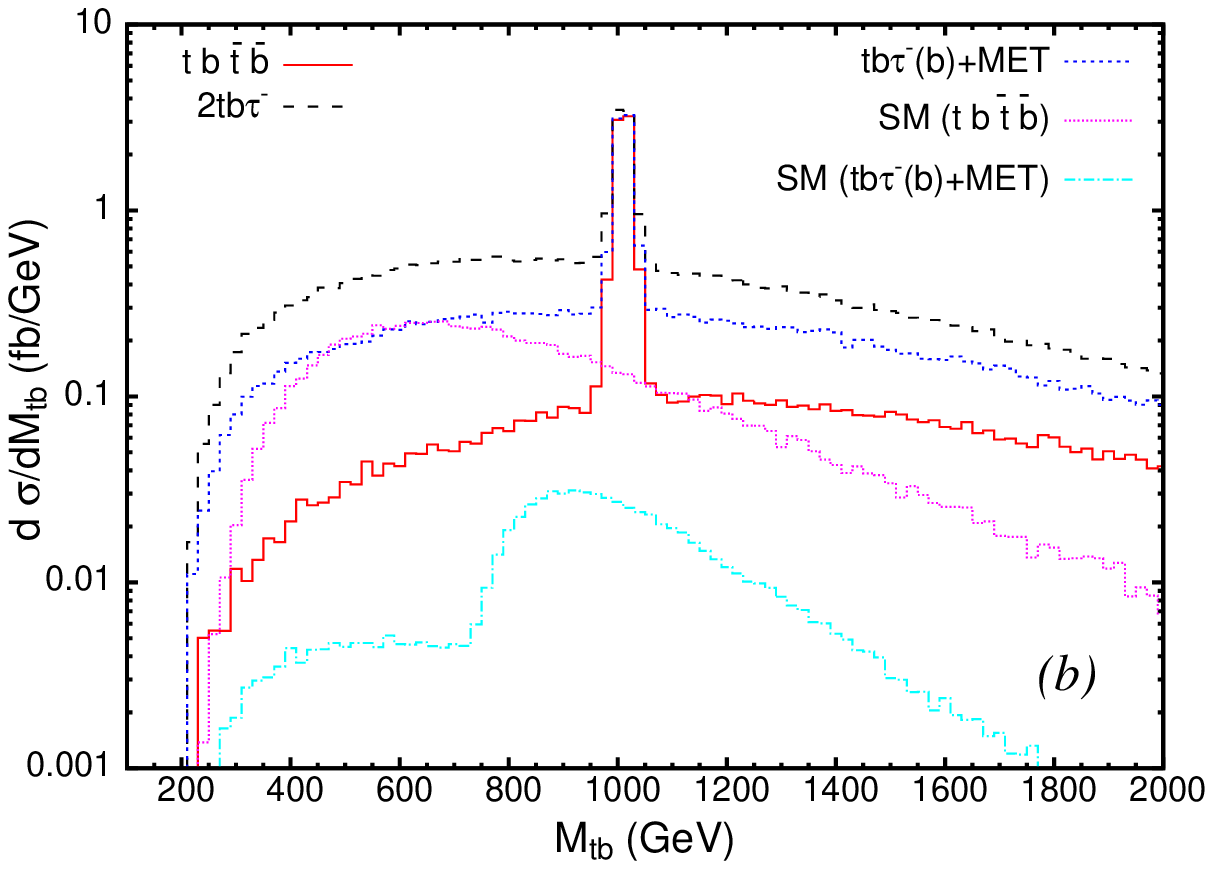}
\caption{\it Invariant mass distribution of $tb$ for the signal and SM background for two 
different choices of leptoquark gauge boson mass, (a) $M_V = 800$ GeV considered at LHC with 
$\sqrt{s}=8$ TeV and (b) $M_V = 1$ TeV considered at LHC with $\sqrt{s}=14$ TeV.}
\label{fig:tb}
\end{figure}
We now look at the final states which correspond to resonant signals for the $Y$ gauge boson. 
We therefore consider the resonance given by {\bf (C3)} and show the invariant mass distribution of 
the top-bottom pair $tb$ in Fig.(\ref{fig:tb}). Note that one of the dominant decay mode for the 
$Y_\mu$ gauge boson gives neutrinos in the final states that leads to large 
missing transverse energy (MET) and is not suitable to reconstruct the $Y_\mu$ mass. However, 
allowing one $Y$ to decay in the neutrino mode still allows reconstruction of 
the other in the visible decay modes of $tb$ and $t\tau$. A large MET in the 
final state also helps in suppressing large contributions to the SM background 
through all hadronic final states which proceed through strong interactions. In Fig.(\ref{fig:tb}) 
we consider four different final state signals which lead to a resonance in the $tb$ invariant 
mass, namely $t\bar{t}bb,~ttb\tau^-,~tb\tau^-\slashed{E}_T$ and $tbb\slashed{E}_T$. The 
signal $ttb\tau^-$ remains the same as discussed for Fig.(\ref{fig:btau}) with the only difference 
being that the contribution coming from the $X\bar{X}$ pair production now acts 
to smear out the resonance in $tb$ invariant mass distribution coming from the $Y_\mu$. This is
the cleanest mode with practically no SM background, although depending on the reconstruction of 
the top quarks. The signal cross section for the $t\bar{t}bb$ final state comes 
from the $Y_\mu$ pair production and for $M_Y=800~(1000)$ GeV at LHC with $\sqrt{s}=8~(14)$ 
TeV is $4.04~(13.75)\times \varepsilon_t^2 ~fb$. The SM background at LHC with $\sqrt{s}=8~(14)$ 
TeV for the signal comes dominantly from three subprocesses with $\sigma(t\bar{t}bb) 
\sim 54.4~(34.1)~fb, ~\sigma(t\bar{t}cc) \sim 55.1~(34.4)~fb$ and $\sigma(t\bar{t}jj) \sim 
10.14~(7.45)~pb$. The stronger cuts at the 14 TeV run is responsible for the 
relatively smaller numbers for the SM background for the higher energy run. Note that after 
including the tagging efficiencies and misstag rates, the corresponding SM background for the 
$t\bar{t}bb$ final state comes out to be $15.18~(9.65)\times \varepsilon_t^2 ~fb$. Although the
SM backgrounds are large in this case, the differential cross section is seen to fall rapidly for 
larger values of the invariant mass. Therefore, a strong cut on the $tb$ invariant mass will be useful
to suppress the background further. For the two final states involving missing transverse energy, we 
have combined their contribution in Fig.(\ref{fig:tb}) under the signal ``$tb\tau^-(b)+MET$". We find that
the SM background for $tbb\slashed{E}_T$ is completely negligible. Note that in the SM background 
for ``$tb\tau^-(b)+MET$", Fig.(\ref{fig:tb}b) shows an unusual kink in the invariant mass distribution of  $M_{tb}$. 
This is a kinematical effect driven by the strong kinematic cuts that we put on the final states. The contribution 
to final state events for $tb\tau^-+MET$ can be isolated into dominant  contributions coming from the on-shell 
production of $t\bar{t}$ and contributions where a $b$-jet is recoiling against a $tW^-$ system 
($2\to 3$ scattering). The strong requirement on the $p_T > 200$ GeV of the final products suppresses the 
$t\bar{t}$ contributions more while affecting the $btW^-$ less which leads to the kink in the invariant mass
distribution. A much weaker requirement of $p_T > 80$ GeV (as is the case with $\sqrt{s}=8$ TeV) or lower 
leads to complete amelioration of the kink like behavior and the SM background distribution looks very 
similar to that in Fig.(\ref{fig:tb}a). The large contribution to the 
background comes from the $\sigma(tb\tau^-\slashed{E}_T) \sim 88.7~(3.05)~fb$ 
at LHC with $\sqrt{s}=8~(14)$ TeV, while the 
$tc\tau^-\slashed{E}_T$ and $tj\tau^-\slashed{E}_T$ are much suppressed due to the small CKM 
mixings between the first two generation quarks and the top quark. 
The SM background after including the efficiency factors is then given as 
$22.16~(0.75) \times \varepsilon_t~fb$, while the signal for $M_Y=800~(1000)$ GeV at the two center
of mass energies is $\sigma(tbb\slashed{E}_T)=4.21~(13.21) \times \varepsilon_t~fb$ and 
$\sigma(tb\tau^-\slashed{E}_T)=4.21~(13.20) \times \varepsilon_t~fb$. Note that $tbb\slashed{E}_T$ 
is the one which gives a resonant signal while $tb\tau^-\slashed{E}_T$ gives a continuum in the $tb$ 
invariant mass distribution because the $t$ and $b$ come from different 
$Y_\mu ~(Y\to \bar{b}\nu_\tau, ~\bar{Y}\to t\tau^-)$. This can be seen in Fig.(\ref{fig:tb}) where the 
large signal contribution in the $tb\tau^-\slashed{E}_T$ channel is spread out in the invariant mass distribution.
Therefore it is instructive to put a $\tau$ veto on the signal with missing transverse momenta when looking
at the invariant mass distribution in $tb$. Again for illustrative purposes we have chosen $\varepsilon_t=1$.    
 
\begin{figure}[t!]
\centering
\includegraphics[height=2.2in]{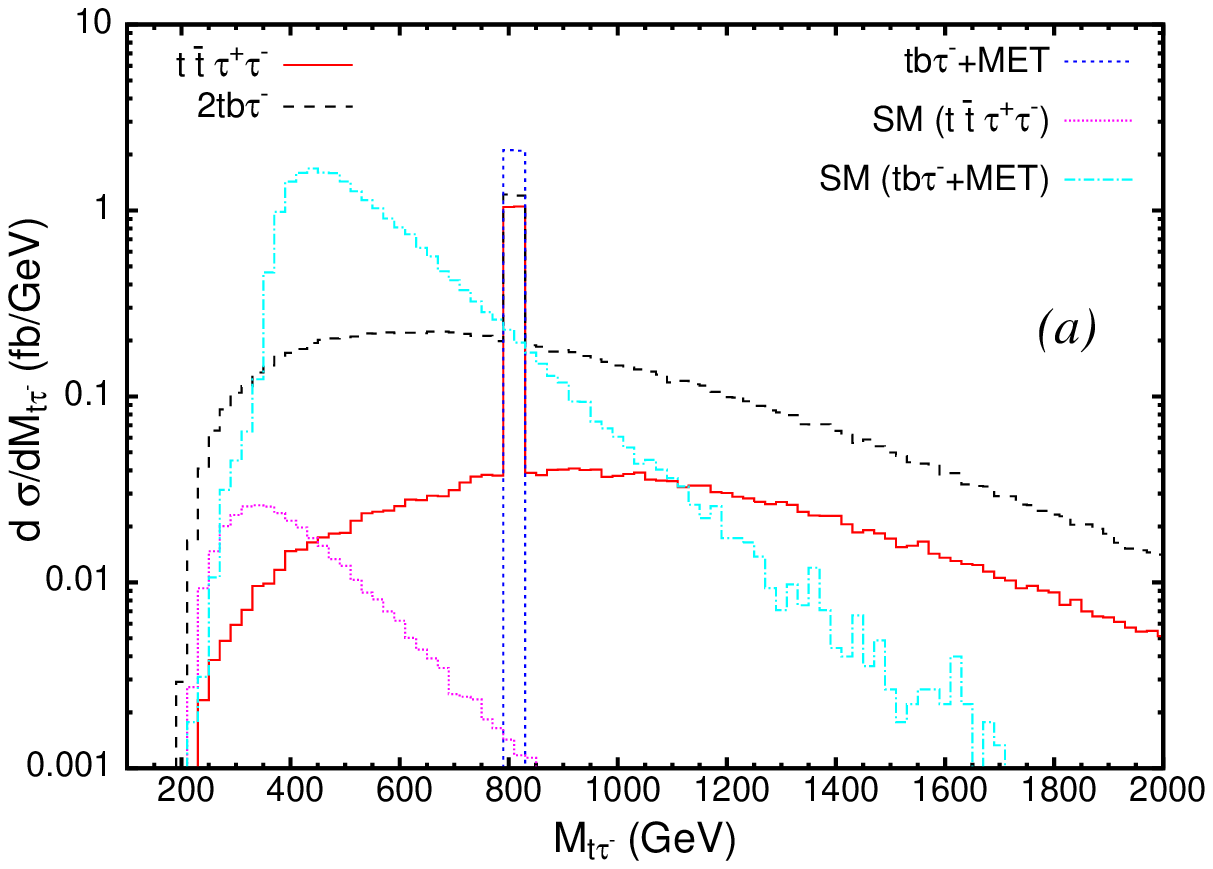}
\includegraphics[height=2.2in]{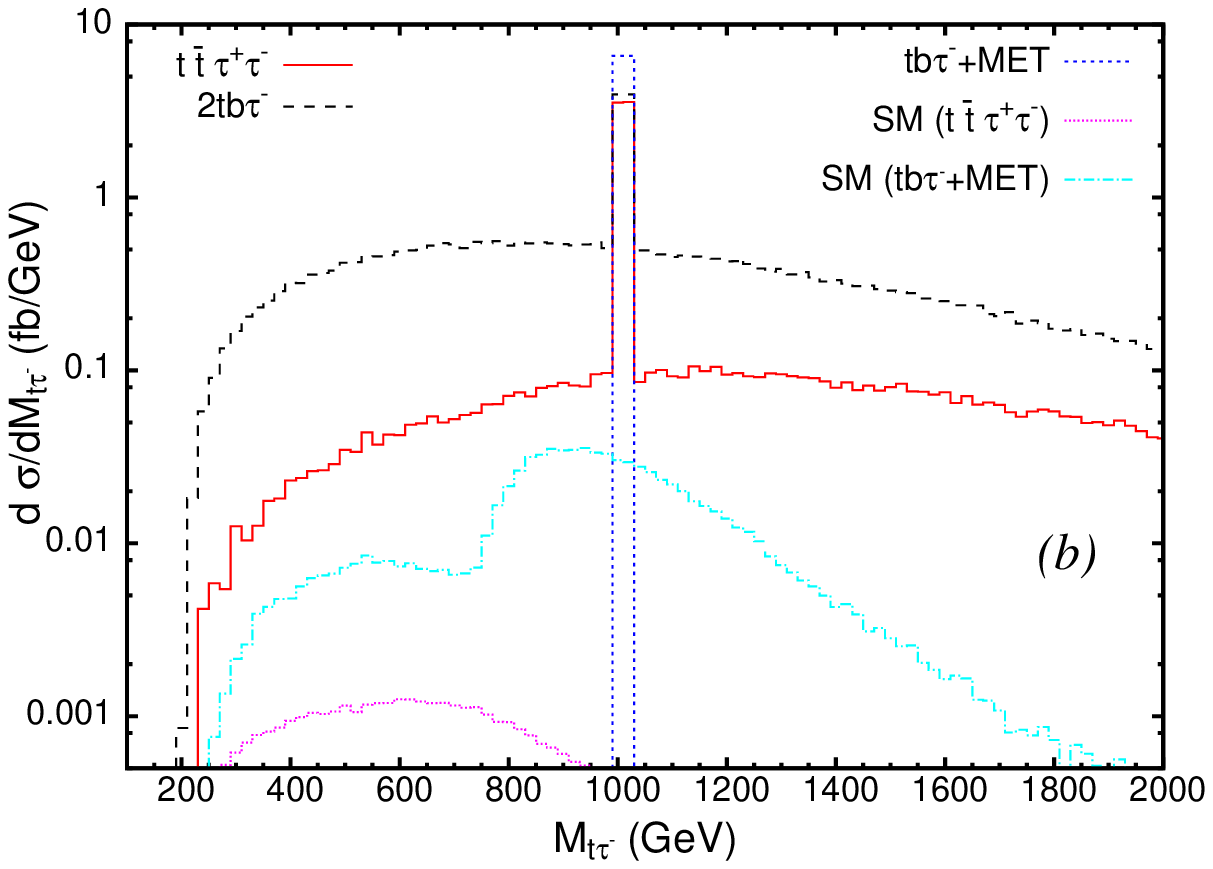}
\caption{\it Invariant mass distribution of $t\tau^-$ for the signal and SM background for two 
different choices of leptoquark gauge boson mass, (a) $M_V = 800$ GeV considered at LHC with $\sqrt{s}=8$ 
TeV and (b) $M_V = 1$ TeV considered at LHC with $\sqrt{s}=14$ TeV.}
\label{fig:ttau}
\end{figure}

We finally consider the resonance given by {\bf (C4)} which again is essential in measuring the charge of the 
$Y_\mu$ gauge boson. To measure the charge one requires the charge measurement of the $\tau$ lepton as
well as the reconstruction of the top quark in its semileptonic channel. We therefore show the invariant mass
distribution in the reconstructed top quark and charged tau lepton pair $(t\tau^-)$ in Fig.(\ref{fig:ttau}) which
corresponds to a resonance for the charge conjugated field of $Y_\mu$.  
The signal is obtained from three different set of final states given by 
$t\bar{t}\tau^+\tau^-, ~ tb\tau^-\slashed{E}_T$ and $2tb\tau^-$. As discussed before the  $2tb\tau^-$ 
contribution is found to have negligible SM background but the contribution from the $X\bar{X}$ 
production to the $t\tau^-$ invariant mass distribution itself acts as a background for the resonant signal from
the $Y\bar{Y}$ production. The $t\bar{t}\tau^+\tau^-$ signal comes solely from the $Y_\mu$ pair production 
and we find that with the proper charge identification of the $\tau$ leptons, we can ignore contributions from
SM background processes such as $t\bar{t}jj$. The signal cross section in this mode is found to be 
$4.04~(13.81) \times \varepsilon_t^2~fb$ for $M_Y=800~(1000)$ GeV at LHC with $\sqrt{s}=8~(14)$ TeV.
The SM background is quite suppressed at both center of mass energy values, given by 
$0.35~(0.04) \times \varepsilon_t^2~fb$. The $tb\tau^-\slashed{E}_T$ signal discussed for the $tb$
resonance in Fig.(\ref{fig:tb}) was found to give a continuum distribution in the $tb$ invariant mass.  However 
it leads to a resonance in the $t\tau^-$ invariant mass distribution as $\bar{Y}\to t\tau^-$. The event rates
are the same as before but one can clearly see a distinct resonance confined to a few bins in the 
invariant mass distribution of $t\tau^-$ in Fig.(\ref{fig:ttau}) for the $tb\tau^-\slashed{E}_T$ signal. 
The large SM background for this mode can again be suppressed with a significantly strong cut on the 
$t\tau^-$ invariant mass. Note again that for the SM background for the $M_{t\tau^-}$  invariant 
mass distribution at $\sqrt{s}=14$ TeV, a similar kink like feature is observed in Fig.(\ref{fig:ttau}b). This 
is because the same subprocess which contributes in Fig.(\ref{fig:tb}b) also features in this case with similar kinematic cuts which we have already discussed before. We must however point out that if strong 
$p_T$ requirements for the final products were put on events for $\sqrt{s}=8$ TeV, we get a similar kink 
like behavior in the invariant mass distribution.

\subsection{LHC sensitivity to the $X_\mu$ and $Y_\mu$ gauge bosons}
As evident from our analyses of the resonant signals for the $X_\mu$ and $Y_\mu$ gauge 
bosons in our models, the LHC would be able to see the signals in various different channels for 
significantly large values of their mass.
A single channel analysis in the $b\tau$ mode relevant for $X_\mu$ search was considered for 
its search at the 7 TeV run of LHC \cite{Chakdar:2012kd,Chatrchyan:2012sv} while another 
experimental study relevant for the $Y_\nu$ search in the $bb\slashed{E}_T$ channel has been 
done by the CMS Collaboration \cite{Chatrchyan:2012st}. Here we do a more expansive sensitivity 
reach at the LHC for these gauge bosons that can be obtained at different integrated luminosities. 
For the top decaying semileptonically to $b\ell^+\nu_\ell$ where $\ell=e,\mu$ the events will be 
at most, or less than $\sim 22\%$ of the reconstructed top events. While it would be $\sim 66\%$ in 
the hadronic decay mode. Thus it gives a clear demarcation on the event rate we specify for the final 
states involving the top and anti-top quarks that would lead to any signal events to reconstruct 
the tops. 

For the sensitivity analysis we define the signal to be observable if the lower limit on the
signal plus background is larger than the corresponding upper limit on the 
background \cite{Sayre:2011ed} with statistical fluctuations
\begin{align*}
L (\sigma_s + \sigma_b) - N \sqrt{L (\sigma_s + \sigma_b)} \geq L\sigma_b + N\sqrt{L \sigma_b}
\end{align*}
or equivalently,
\begin{align}
\sigma_s \geq \frac{N}{L} \left[N + 2\sqrt{L\sigma_b}\right],
\label{eq:conflev}
\end{align}
where $L$ is the integrated luminosity, $\sigma_s$ is the signal cross section, and $\sigma_b$ is the
background cross section. The parameter $N$ specifies the level or probability of discovery.
We take $N = 2.5$, which corresponds to a $5\sigma$ signal. For $\sigma_b  \gg \sigma_s$, this 
requirement becomes similar to
\begin{align}
\mathcal{S} = \frac{N_s}{\sqrt{N_b}} = \frac{L\sigma_s}{\sqrt{ L\sigma_b}} \geq 5 ,
\end{align}
where $N_s$ is the number of events for the signal, $N_b$ is the number of events for the background, 
and  $\mathcal{S}$ equals the statistical significance. 

\begin{table}[!t]
\begin{center}
\begin{tabular}{|c|c|c|c|}
\hline 
${\mathcal Final ~State}$ &  $\sigma_{{\mathcal SM}}~(fb)$ & ${\mathcal Final ~State}$
& $\sigma_{{\mathcal SM}}~(fb)$  \\ 
\hline  $2b\tau^+\tau^-$ &   0.12 ~(0.002)
       &  $ttb\tau^-,~\bar{t}\bar{t}b\tau^+$ & -- \\ 
\hline  $tt\bar{t}\bar{t}$ & 2.31~(24.34) 
       &  $t\bar{t}\tau^+\tau^-$ & 0.35~(0.04) \\
\hline  $2b t\bar{t}$ & 15.18~(9.65)
       &  $2b\slashed{E}_T$ & 25.06~(3.83) \\
\hline $b t \tau^- \slashed{E}_T$ &  22.16~(0.75) 
       & $b \bar{t} \tau^+\slashed{E}_T$ &  22.16~(0.75) \\
\hline  $2b t \slashed{E}_T$ & 0.003~(0.001)  
       &  $2b\bar{t}\slashed{E}_T$ &  0.001~(0.0006)\\
\hline
\end{tabular}
\caption{\textit{The combined SM cross sections estimated at parton level using MadGraph 5 for the 
different final state signals at LHC with $\sqrt{s}=8$ TeV and $\sqrt{s}=14$ TeV. The 14 TeV values
are given in parenthesis. Note that the cross sections given satisfy the kinematic cuts listed
in Table \ref{tab:cuts} and all tagging efficiencies and misstag rates are included.} } 
\label{tab:cross sections}
\end{center}
\end{table}
 In Table \ref{tab:cross sections}, we have calculated the SM background for the different final states that 
we have considered for the signal coming form the pair productions of the $X_\mu$ and 
$Y_\mu$ gauge bosons. The cross sections shown in Table \ref{tab:cross sections} are obtained after passing the events through the 
kinematic selection conditions given in Table \ref{tab:cuts}. In most cases the SM 
backgrounds are quite small and would remain negligible even with an integrated luminosity 
of 100 fb$^{-1}$. Note that as the top reconstruction would require sufficient events after it has 
decayed, we need much larger cross sections for the final states involving top 
quarks. To use Eq.(\ref{eq:conflev}), we require the background events to be 
sufficiently large such that the fluctuations to a Gaussian distribution could be applied. We find that the best reaches are obtained 
 for  the  $bb\slashed{E}_T, ~b\bar{t}\tau^+\slashed{E}_T$ and $bt\tau^-\slashed{E}_T$
final states. For the $bb\slashed{E}_T$ final state at LHC with $\sqrt{s}=8$ TeV, the 
signal cross section for a  $5\sigma$ sensitivity must be greater than $8.54,~5.91,~4.78~fb$
for an integrated luminosity of $L=10,~20,~30~fb^{-1}$ respectively. This 
corresponds to the mass reach of $M_Y=737,~772,~793$ GeV respectively. With the 
higher center of mass energy option for LHC with $\sqrt{s}=14$ TeV, the 
signal cross section for a  $5\sigma$ sensitivity must be greater than $1.995,~1.041,~0.586~fb$
for an integrated luminosity of $L=30,~100,~300~fb^{-1}$ respectively. These lead 
to a mass reach of $1325,~1440,~1545$ GeV respectively. For the  other channels involving the
top quark in the final state, we assume the reconstruction efficiency for the top quark 
$\varepsilon_t \simeq 0.5$ which includes the event loss from kinematic cuts after the top decays.
Adding the contributions for $b\bar{t}\tau^+\slashed{E}_T$ and $bt\tau^-\slashed{E}_T$ 
we find that at the 8 TeV run of LHC, the mass reach is $770,~795$ GeV for an 
integrated luminosity of $L=20,~30~fb^{-1}$ respectively while at the 14 TeV run 
of LHC, where we use the high luminosity options of $200~fb^{-1}$ and 
$300~fb^{-1}$, the $5\sigma$ sensitivity comes out to be about $1650$ GeV and $1690$ GeV respectively. 




\section{Summary and Conclusions}

Although the Standard Model, based on local gauge symmetries, accidentally conserve baryon and lepton numbers, there is no fundamental reason for the baryon and lepton numbers to be exact symmetries of Nature. In fact, Grand Unification, unifying quarks and leptons, naturally violate baryon and lepton number. The remarkable stability of the proton dictate that the masses of these leptoquark and diquark gauge bosons to be at the $10^{16}$ GeV scale. However, baryon and lepton number violating interaction involving only the 3rd family of fermions is not much constrained experimentally. Inspired by the topcolor, topflavor and top hypercharge models, we have a top-GUT model where only the third family of fermions are unified in an $SU(5)$ with the symmetry breaking scale at the TeV. These models give baryon and lepton number violating gauge interactions which involve only the third family, and with interesting resonant signals at the LHC. 

We have  proposed two models, the minimal and renormalizable top $SU(5)$  where the 
$SU(5)\times SU(3)'_C \times SU(2)'_L \times U(1)'_Y$ gauge symmetry 
is broken down to the Standard Model (SM)
gauge symmetry via the bifundamental Higgs fields at low energy. The first 
two families of the SM fermions are charged under $ SU(3)'_C \times SU(2)'_L \times U(1)'_Y$
while the third family is charged under $SU(5)$. 
In the minimal top $SU(5)$ model, we showed that the quark CKM mixing
matrix can be generated via dimension-five operators, and the
proton decay problem can be solved by fine-tuning the coefficients of the
high-dimensional operators at the order of $10^{-4}$.
In the renormalizable top $SU(5)$ model, we introduced additional vector-like fermions whose renormalizable interactions with the SM particles generate these dimension 5 interactions and  we can explain the quark CKM mixing matrix by introducing the vector-like particles, and also there is no proton decay problem.
We have discussed the phenomenology of the models in details 
looking for the resonant signals for the baryon and lepton number violating leptoquark as well as diquark gauge bosons at the LHC,
as well as the various final state arising from the productions and decays of these heavy gauge bosons. We have also calculated the corresponding SM backgrounds. We find that a $5\sigma$ signal can be observed for a mass leptoquark / diquark of about 770/800 GeV at the 8 TeV LHC with luminosity of $20 fb^{-1}$/$30 fb^{-1}$ . The mass reach extends to about 1450 TeV for 14 TeV LHC with a luminosity of $100 fb^{-1}$.

\begin{acknowledgments}

This research was supported in part by the Natural Science Foundation of China 
under grant numbers 10821504, 11075194, and 11135003, 
and by the United States Department of Energy Grant Numbers DE-FG03-95-Er-40917, DE-FG02-04ER41306.
The work of S.K.R. was partially supported by funding available from the Department of Atomic Energy, 
Government of India, for the Regional Centre for Accelerator-based Particle Physics, Harish-Chandra Research Institute.

\end{acknowledgments}


\end{document}